\documentclass[bj,authoryear]{imsart}

\usepackage{algorithm}
\usepackage{algpseudocode}
\usepackage{amsmath}
\usepackage{amsfonts}
\usepackage{comment}
\usepackage{tikz}
\usepackage{booktabs}
\usepackage{pdfpages}
\usetikzlibrary{decorations.pathreplacing}
\startlocaldefs
\theoremstyle{plain}

\newtheorem{theorem}{Theorem}[section]
\newtheorem{lemma}[theorem]{Lemma}
\newtheorem{proposition}{Proposition}
\newtheorem{corollary}{Corollary}
\newtheorem{remark}{Remark}
\theoremstyle{definition}
\newtheorem{definition}[theorem]{Definition}
\newtheorem*{example}{Example}

\newcounter{assumptionH}
{%
\medskip
\noindent\textbf{Assumptions.}
  \begin{enumerate}[({H}1)]%
  \setcounter{enumi}{\value{assumptionH}}%
}{%
  \setcounter{assumptionH}{\value{enumi}}%
  \end{enumerate}
}

\renewcommand{\theenumi}{\Alph{enumi}}

\newcounter{assumptionHt}

{%
\begin{enumerate}[({G}1)]%
    \setcounter{enumi}{\value{assumptionHt}}%
    }{%
    \setcounter{assumptionHt}{\value{enumi}}%
\end{enumerate}
}

\newcommand{\EE}{\mathbb{E}}

\newcommand{\PP}{\mathbb{P}}

\newcommand{\Tt}{\boldsymbol t}

\newcommand{\Ss}{\boldsymbol s}

\newcommand{\Rplus}{\mathbb R_+}

\endlocaldefs

\begin{document}

\begin{frontmatter}
\title{Structural adaptation and rate accelerated estimation in   bivariate   functional data}
\runtitle{Structural adaptation in bivariate functional data}

\begin{aug}
\author[A]{\inits{F.}\fnms{Omar}~\snm{Kassi}\ead[label=e1]{omar.kassi@ensai.fr}}
\author[B]{\inits{S.}\fnms{Sunny G. W.}~\snm{Wang}\ead[label=e2]{sunny.wang@ensai.fr}}
\address[A]{Univ Rennes, Ensai, CNRS, CREST-UMR 9194, F-35000 Rennes, France\printead[presep={,\ }]{e1}}

\address[B]{Univ Rennes, Ensai, CNRS, CREST-UMR 9194, F-35000 Rennes, France\printead[presep={,\ }]{e2}}
\end{aug}

\begin{abstract}
We introduce directional regularity, a new definition of anisotropy for multivariate functional data. Instead of taking the conventional view, which determines anisotropy as a notion of smoothness along a dimension, directional regularity additionally views anisotropy through the lens of directions. We show that faster rates of convergence   for smoothing    can be obtained through a change-of-basis by adapting to the 
  anisotropy  of a bivariate process. An algorithm for the estimation and identification of the change-of-basis matrix is constructed, made possible due to the replication structure of functional data. Non-asymptotic bounds are provided for our algorithm, supplemented by numerical evidence from an extensive simulation study.  Finally, a real‑world rainfall measurement dataset is analyzed with our methods.    
\end{abstract}

\begin{keyword}
\kwd{Anisotropy}
\kwd{adaptive algorithms}
\kwd{effective smoothness}
\kwd{multivariate functional data}

\end{keyword}

\end{frontmatter}

\section{Introduction}
Functional data analysis (FDA) is a burgeoning field that deals with modern, complex data, such as those collected by sensors. It has proven successful in a wide range of applications, such as medical, economics, and sports data (see for example, \cite{Ramsay2005}, \cite{Horvath2012}, \cite{Kokoszka2017}). A distinctive feature of FDA is that the data are functions, defined over a domain $\mathcal{T}$, such as curves, surfaces and their $d$-dimensional equivalents. 

When $\mathcal{T}$ is multivariate, e.g., $\mathcal{T} = (0, 1)^2$, detecting anisotropy is important in many functional data applications. For instance, in climatology, identifying the principal direction of anisotropy can improve spatial models \citep{niemi2015, friedland2016, oliveira2022}, whereas in fingerprint recognition, orientation‑dependent variability can enhance matching accuracy \citep{wang2005, greenberg2002, somvanshi2012}.

Moreover, since functional observations are invariably recorded on a discrete grid, practitioners routinely apply smoothing methods, such as splines, kernel smoothers, or basis expansions, as a preprocessing step. It is well known that the optimal smoothing parameters depend on the underlying regularity of the process, and hence on any anisotropy present.

To informally motivate our contributions, (formalized in Section \ref{sec:appli}), consider a real-valued, bi-variate, second order process $\left\{X(\mathbf{t}),  \mathbf{t} \in (0,1)^2 \right\}$ with regularity $H_i$   (see Definition \ref{def:direg-def})   along the direction $\mathbf{u}_i \in \mathbb{S}$  (denoting the unit circle),  where $|H_i - H_j| > 0$   for $i,j \in \{1,2\}$ with   $j \neq i$. We observe noisy samples $(Y_m^{(j)}, \mathbf{t}_m) \in \mathbb{R} \times (0,1)^2$, generated under the model
\begin{equation}\label{eq:data-model}
    Y_m^{(j)} = X^{(j)}(\mathbf t_m) + \varepsilon_m^{(j)}, \qquad 1 \leq m \leq M_0, 1 \leq j \leq N,
\end{equation}
where $\varepsilon^{(j)}_m$ are  i.i.d (in $m$ and $j$) centered errors with constant variance $\sigma^2$.   The noise framework considered in model \eqref{eq:data-model} is common in many applications where the measurement errors are naturally not correlated across subjects and time. Examples include CD4 cell counts \citep{Yao2005} and human gait data \citep{rice2001}.
  
Suppose the goal is to obtain an estimate $\widehat X(\mathbf{t}, h_1, h_2)$ from the observed data. Using a linear smoother, e.g., the Nadaraya-Watson estimator or smoothing splines, one can show under suitable assumptions that the optimal smoothing parameters satisfy
\begin{equation*}
    h_i^* \asymp M_0^{-\frac{H_j}{2H_iH_j + H_i + H_j}}, \qquad i \neq j,
\end{equation*}
where the exponent reflects the impact of anisotropy. In particular, bandwidths should be adapted to the varying regularities along different directions. Since data are typically observed in the canonical basis, bandwidth selection is often based on regularities observed in that basis representation. However, this approach can be suboptimal, especially if anisotropy is present only on another basis. It is therefore important for practitioners to determine whether anisotropy exists, and if so, to identify the directions in which it occurs along with their corresponding regularities. 

  To illustrate how adapting the bandwidth to anisotropy can accelerate the rate of convergence, suppose that $ \{\mathbf u_1, \mathbf u_2\} \ne \{\mathbf e_1, \mathbf e_2\}$, where $\{\mathbf e_1, \mathbf e_2\}$ is the canonical basis. In this case, one can show that the regularity along $\mathbf e_1$ and $\mathbf e_2$ is $\underline H:=\min\{H_1,H_2\}$.
On the canonical basis, this leads to a theoretically optimal bandwidth choice of $h_i \asymp M_0^{-1/(2\underline H +2)}$, yielding the rate of convergence $ M_0^{-\underline H/(2\underline H +2)}
$ for isotropic smoothing. This rate is strictly slower than the rate of convergence $M_0^{- H_1H_2/(2 H_1H_2 +H_1+H_2)}$ in the case of anisotropic regression. This motivates the identification of $ \{ \mathbf u_1, \mathbf u_2\}$ in accelerating convergence rates.

The contributions of this paper include (i) a characterization of how regularity properties may vary with orientation (Section \ref{sec:direg}); (ii) the development of novel algorithms to detect the presence and direction of anisotropy, supported by theoretical guarantees (Section \ref{sec:methodo});   
(iii) a numerical evaluation of the proposed methods (Section \ref{sec:numerics}), together with an application to real anisotropic data; 
(iv) a demonstration of how these algorithms can improve the statistical analysis of bivariate functional data through a change-of-basis (Section \ref{sec:appli}); (v) a discussion on other important aspects of the algorithms with possible extensions (Section \ref{sec:discussion}).

\subsection{Related work}
Considering and adapting to the directional nature of anisotropy has emerged in a relatively recent body of work in statistical learning, called structural adaptation \citep{lepski-aniso-2015}. See \citep{samarov-tsybakov-2004, lepski-gilles-2020,ammous-2024-dir}  for some examples in the density model. In spatial data, this has been considered under the umbrella of fractal analysis, for example in the seminal work of \citep{davies1999fractal}. Identifying the spatial anisotropy is a natural goal in several applications such as climatology, with an example being rainfall data \citep{bernardi2018}.

We build upon a recent body of work in FDA attempting to characterize the regularity of a stochastic process $X$ through the approximate identity
\begin{equation*}
\mathbb{E}[\{X(u) - X(v) \}^2] \approx L^2_t |u - v|^{2H_t}, \qquad \forall u \leq t \leq v, \quad t \in \mathcal{T} \subset \mathbb{R},
\end{equation*}
where the regularity is defined as the first-order exponent $H_t$ associated with an infinitesimal increment. See for example \citep{Golovkine2021,wang2023adaptive,kassi2023,maissoro2024}. We extend these ideas and consider anisotropy in FDA by taking into account the direction of this infinitesimal increment.

\begin{figure}[h]
\caption{  Let $H_2 > H_1 > 0$ be the regularity along the direction $\mathbf{u}_1$ and $\mathbf{u}_2$ respectively.     In general, the worst regularity $H_1$ will be obtained on the canonical basis, leading to isotropy. This can lead to slower rates of convergence. A change-of-basis from $(\mathbf{e}_1, \mathbf{e}_2)$ to $(\mathbf{u}_1, \mathbf{u}_2)$ enables one to obtain an anisotropic process. Locating the basis ($\mathbf{u}_1, \mathbf{u}_2$) is equivalent to locating the angle between $\mathbf{u}_1$ and $\mathbf{e}_1$. } 
\label{fig:direg_plot}
\includegraphics[scale=0.6]{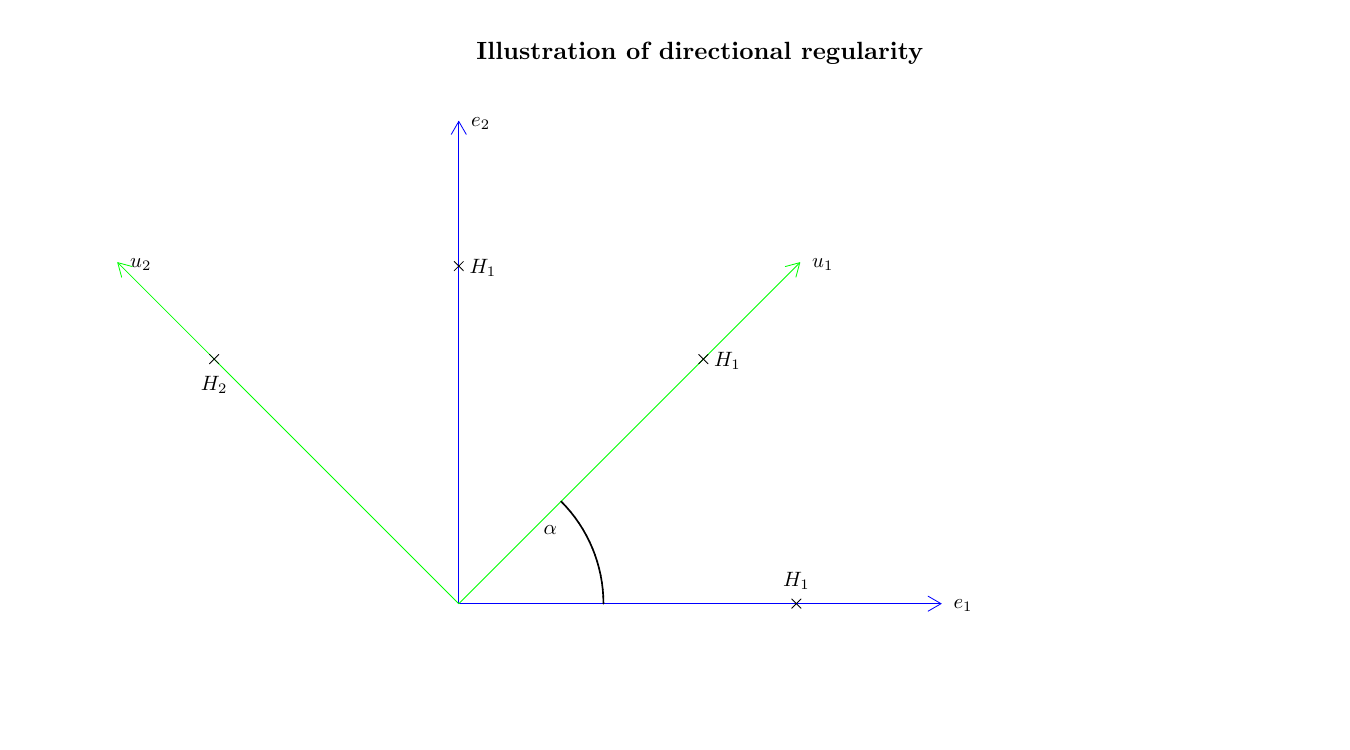}
\end{figure}

\section{Directional regularity}\label{sec:direg}
\subsection{Data setting and definition}
Let $\{X(\mathbf{t}), \mathbf{t} \in \mathcal{T} \}$ be a second-order stochastic process on an open, bounded domain $\mathcal{T} \subset \mathbb{R}^2$. In this paper, we will focus on the common design framework,   where observations are generated according to model \eqref{eq:data-model},   which represents many applications. 
In particular, functional data collected by sensors and images fall into such a framework. We suppose without loss of generality that the design points $\mathbf{t}_m$ are observed on the canonical basis, and that $\mathcal{T} = (0, 1)^2$. The random design case is discussed in Section \ref{sec:discussion}.   Unless otherwise stated, all the vectors below are column vectors, and for vectors $\mathbf{u}, \mathbf{v} \in \mathbb{R}^2$ , $\mathbf{u}^{\top}$  denotes the transpose of $\mathbf{u}$, and $\mathbf{u}^\top\mathbf{v}$ denotes the scalar product in $\mathbb{R}^2$.

\begin{definition}\label{def:direg-def}
Let $X$ be a stochastic process with  continuous and non-differentiable sample paths, and $\mathbf{u} \in \mathbb S$. The process $X$ has local regularity $H_{\mathbf u}$ at $\mathbf{t}\in \mathcal T$ along the direction $\mathbf{u}$ if a non-negative function  $L_u : \mathcal T \rightarrow \mathbb{R}_+$ exists such that  : 
\begin{equation}\label{eq:mean-squared-variation}
\theta_{\mathbf u}(\mathbf{t}, \Delta):=\mathbb{E}\left[\left\{  X\left( \mathbf{t} - \frac{\Delta}{2} \mathbf{u} \right)- X\left( \mathbf{t} + \frac{\Delta}{2}\mathbf{u} \right) \right\}^2 \right] = L_{\mathbf{u}}(\mathbf{t}) \Delta^{2H_{\mathbf{u}}(\mathbf{t})} + G_{  \mathbf{u}   } (\mathbf{t},\Delta), 
\end{equation}
where $G_{\mathbf{u}} (\mathbf{t}, \Delta) = o\left(\Delta^{2H_{\mathbf{u}}(\mathbf{t})} \right)$, as $\Delta \rightarrow 0$, for each fixed $\mathbf t$. The map
\begin{equation*}
H: 
\begin{cases}
\mathbb{S} \times \mathcal{T} \rightarrow (0, 1) \\
(\mathbf{u}, \mathbf{t}) \mapsto H_{\mathbf{u}}(\mathbf{t})
\end{cases}
\end{equation*}
is called the directional regularity of $X$.
\end{definition}

Definition \ref{def:direg-def} is local in the sense that the local regularity may vary along the domain $\mathcal{T}$. In the following, we consider the case where $G_{\mathbf{u}}(\mathbf{t}, \Delta) = O(\Delta^{2H_{\mathbf{u}}(\mathbf{t}) + \zeta})$ for some $\zeta>0$. If the function $H$ does not depend on the direction $\mathbf u$, we say that $X$ is an \textit{isotropic} process, otherwise we call it an \textit{anisotropic} process. Definition \ref{def:direg-def} formalizes anisotropy as a notion of regularity not only along a dimension, but also along a  direction. This is in contrast to the usual consideration of anisotropy, which is typically determined by examining the regularity on the canonical basis.  See for example, \cite{bertin2004}, or Definition 1 in \cite{amorino2021}.   Some examples of processes with prescribed directional regularity will be discussed in Section \ref{sec:examples}. 

A natural question that arises is the number of possible regularities that an anisotropic process can take, and how these regularities change with directions. Lemma \ref{lm} provides a precise characterization to the preceding question, stating that there can be at most two different regularities. Furthermore, there exists a unique direction, up to a reflection, for the maximal regularity.

\begin{lemma}\label{lm}
Let $\mathbf t \in \mathcal T$ and assume that there exist basis vectors $(\mathbf{u}_1(\mathbf t), \mathbf{u}_2(\mathbf t)) \in  \mathbb{S}\times \mathbb{S}  $ such that $H_{\mathbf{u}_1}(\mathbf{t}) < H_{\mathbf{u}_2}(\mathbf{t})$. Moreover, suppose that the functions $ L_{\mathbf{u}_1}$ and $L_{\mathbf{u}_2}$  are continuously differentiable. For any $\mathbf{v} \in \mathbb{S}$, we have the following dichotomous relationship:
\begin{itemize}
\item If $\mathbf{v}\ne \pm \mathbf{u}_2$, then $H_{\mathbf{v}}(\mathbf{t}) = H_{\mathbf{u}_1}(\mathbf{t})$.

\item Otherwise, we have $H_{\mathbf{v}}(\mathbf{t}) = H_{\mathbf{u}_2}(\mathbf{t})$.
\end{itemize}
\end{lemma}
\begin{proof}
Fix $\mathbf t \in \mathcal T$. When $\mathbf{v} = \pm \mathbf{u}_2$, Lemma \ref{lm} is a direct consequence of Definition \ref{def:direg-def}. For $\mathbf{v} \neq \pm \mathbf{u}_2$, Lemma \ref{lm} can be established by examining the mean-squared variation $\theta_{\mathbf{v}}$. It is shown in the Supplementary Material \citep{kassiwangsupp25} that 
\begin{multline}\label{eq:theta-v}
    \theta_{\mathbf{v}}(\mathbf{t}, \Delta)= L_{\mathbf{u}_1}(\mathbf{t}) | \mathbf{v}^\top \mathbf{u}_1  \Delta|^{2H_{\mathbf{u}_1}  (\mathbf t)   }+L_{\mathbf{u}_2} (\mathbf{t})| \mathbf{v}^\top \mathbf{u}_2  \Delta|^{2H_{\mathbf{u}_2}  (\mathbf t)   }+O\left(\Delta^{H_{\mathbf{u}_1}(\mathbf t)+ H_{\mathbf{u}_2}(\mathbf t)} \right) \\
+ O \left(\Delta^{2H_{\mathbf{u}_1}  (\mathbf t)   +\zeta} \right)+O \left(\Delta^{2H_{\mathbf{u}_1}(\mathbf t)+1}\log(\Delta) \right)O \left(\Delta^{2H_{\mathbf{u}_2}  (\mathbf t)   +\zeta} \right)+O \left(\Delta^{2H_{\mathbf{u}_2}(\mathbf t)+1}\log(\Delta) \right).
\end{multline}
When $H_{\mathbf{u}_1}(\mathbf t) < H_{\mathbf{u}_2}(\mathbf t)$, the dominating term in \eqref{eq:theta-v} is given by   $L_{\mathbf{u}_1}(\mathbf{t}) |\mathbf{v}^\top \mathbf{u}_1\Delta|^{2H_{\mathbf{u}_1}  (\mathbf t)   }$    for $\Delta$ sufficiently small. By Definition \eqref{def:direg-def}, the regularity at $\mathbf t$ is given by $H_{\mathbf{u}_1}(\mathbf t)$. The converse holds when $H_{\mathbf{u}_1} (\mathbf t)> H_{\mathbf{u}_2}(\mathbf t)$.
\end{proof}

Lemma \ref{lm} states that the map $\mathbf v \mapsto H_{\mathbf v}$ can take at most two possible values. This is in alignment with the results in \citep{davies1999fractal}, who studied anisotropy from the lens of the fractal dimension of surfaces, and showed that "the fractal dimension of line transects
across a surface must either be constant in every direction or be constant in each direction except
one".

 For a fixed $\mathbf t \in \mathcal T$  , let $\mathbf{v}^* (\mathbf t)= {\arg\max}_{\mathbf v \in \mathbb S} H_{\mathbf v}(\mathbf{t})$ be the direction that maximizes the regularity of $X$ in the sense of Definition \ref{def:direg-def}. Then $\mathbf{v}^*(\mathbf t)$ and $-\mathbf{v}^*(\mathbf t)$ are the only two possible maximizers, and are "singularities" in the sense that if we locally examine any vector other than $\pm \mathbf{v}^*(\mathbf t)$, the associated regularity is $\min_{u \in \mathbb{S}} H_{\mathbf{u}}(\mathbf{t})$. Parameterizing the vectors by their angle $\mathbf u (\beta)= \cos(\beta) \mathbf e_1 + \sin(\beta)\mathbf e_2 $, the maximization problem $(\mathcal{P}) : \arg\max_{\mathbf{v} \in \mathbb{S}} H_{\mathbf v}(\mathbf{t})$  is equivalent to ${\arg\max }_{\beta \in [0,2\pi)} H_{\mathbf u (\beta)}(\mathbf{t})$. By restricting the domain to $[0, \pi)$, $(\mathcal{P})$ admits a unique solution. $(\mathcal{P})$ can thus be parameterized as locating the angle between $\mathbf{v}^*$ and $\mathbf{e}_1$, given by
\begin{equation}\label{maximal_angle}
\alpha(\mathbf t)= {\arg\max }_{\beta \in [0,\pi)} H_{\mathbf u (\beta)}(\mathbf{t}).
\end{equation}
An illustration can be found in Figure \ref{fig:direg_plot}.


\subsection{Examples}\label{sec:examples}
In the following, we provide some examples of processes that motivate Definition \ref{def:direg-def}.

\begin{example}[Sums and products of two fractional Brownian motions]
Let $H\in (0,1)$ be the Hurst parameter of the fractional Brownian motion (fBm) $\{B^H(t), t\in (0,1) \}$. $B^H$ is a centered Gaussian process with covariance function
\begin{equation*}
\mathbb E\left[ B^H(t)B^H(s)\right]= \frac{1}{2}\left\{ |t|^{2H}+ |s|^{2H} - |t-s|^{2H} \right\}, \qquad \forall (t,s)\in (0,1)^2.
\end{equation*}
The fBm is a generalization of the standard Brownian motion, which has Hurst parameter $H = 1/2$. Unlike the standard Brownian motion, the increments of the fBm are not necessarily independent. A simple calculation reveals that 
\begin{equation}\label{ms:brow}
\mathbb E \left[\{B^{H}(t) - B^H(s)\}^2 \right] = |t-s|^{2H} , \qquad \forall (t,s) \in (0,1)^2.
\end{equation}
Let $(\mathbf{u}_1, \mathbf{u}_2)$ be   an orthonormal    basis of $\mathbb R^2$, $(H_1, H_2)\in (0,1)^2$, and $B_1$,$B_2$ be two independent fBms with Hurst parameters $H_1$ and $H_2$ respectively. Set 
\begin{equation}\label{sum:brows}
    X_1(\mathbf t)= B_1(t_1)+ B_{  2   }(t_2), \qquad \text{ for any }\mathbf t\in (0,1)^2, \quad \text{and } \mathbf t= t_1 \mathbf{u}_1 + t_2 \mathbf{u}_2.
\end{equation}
In view of \eqref{ms:brow}, the independence of $B_1$ and $B_2$ implies that for any $\Delta>0$ sufficiently small, we have for   any vector $\mathbf{v} \in \mathbb{S}$, 
\begin{equation}\label{eq:other-direg}
    \mathbb{E}\left[\left\{X_1\left(\mathbf{t} - \frac{\Delta}{2} \mathbf{v} \right) - X_1\left(\mathbf{t} + \frac{\Delta}{2} \mathbf{v}\right) \right\}^2\right] = \left(\mathbf v^\top \mathbf{u}_1 \right)^{2H_1} \Delta^{2H_1}+\left(\mathbf v^\top \mathbf{u}_2 \right)^{2H_2} \Delta^{2H_2}.    
\end{equation}

By Definition \ref{def:direg-def}, for $i=1,2$, the process $X_1$ has a local regularity $H_i$ along the direction $\mathbf{u}_i$ with $G(\mathbf t, \Delta )=0$ $(\zeta= \infty)$, and $L_{\mathbf{u}_i} \equiv 1$. 
Equation \eqref{eq:other-direg} states that if $H_1=H_2$, $X_1$ has a local regularity $H_1$ along $\mathbf v$ with $L_{\mathbf v}\equiv  (\mathbf v^\top \mathbf{u}_1)^{2H_1} + (\mathbf v^\top \mathbf{u}_2)^{2H_1}$. On the other hand, if $H_1\ne H_2$, say $H_1<H_2$, 
$X_1$ has a local regularity $H_1$ along $\mathbf v \ne \mathbf{u}_2$ with $L_{\mathbf v}\equiv  (\mathbf v^\top \mathbf{u}_1)^{2H_1}$ and $G( \mathbf t, \Delta) = O(\Delta^{2H_1 + 2(H_2-H_1)})$. Thus the only direction which gives the maximum regularity is $\mathbf{v} = \mathbf{u}_2$, and any other direction yields the minimum regularity.

Similarly, we can define the tensor product of $B_1$ and $B_2$ with respect to the basis $(\mathbf{u}_1, \mathbf{u}_2)$:
\begin{equation}\label{product:brows}
X_2(\mathbf t)= B_1(t_1)\times B_{  2  }(t_2), \qquad \text{ for any }\mathbf t\in (0,1)^2, \quad \text{and } \mathbf t= t_1 \mathbf{u}_1 + t_2 \mathbf{u}_2.
\end{equation}
For any $\Delta > 0$ sufficiently small, we analogously have
 
\begin{equation}\label{equa9}
\mathbb{E}\left[\!\left\{X_2\left(\mathbf{t} - \frac{\Delta}{2} \mathbf{v} \right) - X_2\left(\mathbf{t} + \frac{\Delta}{2} \mathbf{v}\right) \right\}^2\right]\! =\!  \left\{t_2^{2H_2}\! \left(\frac{\mathbf v ^\top \mathbf{u}_1}{2}\right)^{2H_1} \Delta^{2H_1} + t_1^{2H_1}\left(\frac{\mathbf v ^\top \mathbf{u}_2}{2}\right)^{2H_2} \!\!\Delta^{2H_2}\!\right\}\times \{1+O(\Delta)\},
\end{equation}
so the process $X_2$ has a local regularity $H_i$ along the direction $\mathbf{u}_i$. Similarly, if $H_1 < H_2$, the only direction with the maximal regularity is given by $\mathbf{v} = \mathbf{u}_2$. For any $\mathbf{v} \neq \mathbf{u}_2$, the regularity is given by $H_1$.
  
\end{example}

\begin{example}[Sums
 and products of two independent Ornstein-Uhlenbeck processes]

The stationary fractional Ornstein-Uhlenbeck process $\{ U(t), t\in (0,1)\}$ with index $\rho \in (0,2)$ is a centered Gaussian process, with a covariance function given by 
\begin{equation*}
\mathbb E\left[U(t)U(s) \right]= \exp\left( - a|t-s|^\rho \right), \quad\text{ for some } a>0\text{ and } t,s \in (0,1).
\end{equation*}
The covariance structure exhibits the following property:
\begin{equation*}
\mathbb E \left[\{ U(t)- U(s)\}^2 \right]= 2a |t-s|^\rho + O(|t-s|^{2\rho}), \qquad \text{for any } t,s\in (0,1).
\end{equation*}
Its bi-dimensional counterpart can be constructed by considering either the sum or the tensor product of two independent Ornstein-Uhlenbeck processes, similar to the fBm in \eqref{sum:brows} and \eqref{product:brows}.
\end{example}

\begin{example}[Multi-fractional Brownian sheet]
The local regularity in  previous examples is constant along the domain $\mathcal{T}$. The multifractional Brownian sheet (MfBs) is a generalization of the standard fractional Brownian sheet, where the Hurst parameter is allowed to vary along the domain. See \citep{herbin_06} among others. See \cite[Proposition 6]{kassi2023} for an illustration of the directional regularity property in the case of MfBs. 
\end{example}

\section{Methodology}\label{sec:methodo}
\subsection{Estimating equations}\label{sec:alpha-estim}
  Let $\mathbf t \in \mathcal T$, $(\mathbf{e}_1, \mathbf{e}_2)$ be the canonical basis of $\mathcal{T}$, and $(\mathbf{u}_1, \mathbf{u}_2)$ be orthonormal basis vectors such that $|H_{\mathbf{u}_1}(\mathbf{t}) - H_{\mathbf{u}_2}(\mathbf{t})|>0$.   It is worth noting that $(\mathbf{u}_1, \mathbf{u}_2)$ depends on $\mathbf{t}$, but for simplicity, we drop this notational dependence.   Let $\alpha(\mathbf{t}) \in [0, \pi)$ be the solution of the problem \eqref{maximal_angle}. Denote $H_1(\mathbf{t}) = H_{\mathbf{u}_1}(\mathbf{t})$ and $H_2(\mathbf{t}) = H_{\mathbf{u}_2}(\mathbf{t})$. The estimating equation for $\alpha$ is found in Proposition \ref{prop:alpha-prop}. Denote $a \wedge b = \min\{a, b\}$, for $a, b \in \mathbb{R}$, and $a \vee b = \max\{a, b\}$. Let $\mathbb{I}$ be the indicator function, and $\# \mathcal{S}$ denote the cardinality of a finite set $\mathcal{S}$.  
Recall that, in this paper we consider processes for which $G(\mathbf t, \Delta)=O( \Delta^{H_{\mathbf u} + \zeta})$ for some $\zeta>0$, where $\mathbf u$ is a unit vector and $G_{\mathbf u}(\cdot, \Delta)$ is given in Definition \ref{def:direg-def}.

\begin{proposition}\label{prop:alpha-prop}
Suppose that $\mathbf{u}_1 \neq \pm \mathbf{e}_i$, for $i = 1, 2$. If a process $X$ satisfies \eqref{eq:mean-squared-variation} for any $\mathbf{t} \in \mathcal{T}$, then we have
\begin{equation}\label{eq:g-alpha-true}
 \left(\frac{\theta_{\mathbf{e}_2}(\mathbf{t}, \Delta)}{\theta_{\mathbf{e}_1}(\mathbf{t}, \Delta)} \right)^{ \frac{1}{2\underline{H}(\mathbf{t})}}=g\left(\alpha(\mathbf{t}), \mathbf t \right) + O\left(\Delta^{\zeta \wedge |H_1(\mathbf{t}) -H_2(\mathbf{t})|}  \right),
\end{equation}
where $g(\cdot,\mathbf{t}) = |\tan(\cdot)| \mathbb{I}_{\{H_1(\mathbf{t}) < H_2(\mathbf{t})\}} + |\cot(\cdot)| \mathbb{I}_{\{H_1(\mathbf{t}) > H_2(\mathbf{t}) \}}$, and $\underline{H}(\mathbf{t}) =  H_1(\mathbf{t}) \wedge H_2(\mathbf{t})$.
\end{proposition}
\begin{proof}
    Let $\mathbf t \in \mathcal T$ and set $\widetilde \alpha (\mathbf{t})$ to be the angle between $\mathbf{u}_1$ and $\mathbf{e}_1$. Then $\alpha (\mathbf{t})  \equiv \widetilde \alpha[2\pi]$ or $\alpha \equiv \widetilde \alpha (\mathbf{t})  + \pi/2 [2\pi]$; where the equivalence holds modulo a $2\pi$ translation. Since $ (\mathbf{u}_1, \mathbf{u}_2)$ and $(\mathbf{e}_1, \mathbf{e}_2)$ are orthonormal bases of $\mathbb{R}^2$, we have 
\begin{equation*}
|\mathbf{e}_i^{\top}\mathbf{u}_i| = |\cos(\widetilde \alpha (\mathbf{t}) )|, \quad |\mathbf{e}_i^{\top}\mathbf{u}_j| = |\sin(\widetilde \alpha (\mathbf{t}) )|\quad    \text{ for } i,j=1,2 \text{ and } j \neq i.
\end{equation*}
In view of \eqref{eq:theta-v}, we obtain
\begin{equation*}
\theta_{\mathbf{e}_1}(\mathbf{t}, \Delta)= L_{\mathbf{u}_1} (\mathbf{t}) |\cos(\widetilde \alpha (\mathbf{t}) )\Delta|^{2H_1(\mathbf t)}+ L_{\mathbf{u}_2} (\mathbf{t})|\sin(\widetilde \alpha (\mathbf{t}) )\Delta|^{2H_2(\mathbf t)} + O \left(\Delta^{2\underline H(\mathbf t)+\{\zeta\wedge |H_1(\mathbf t)- H_2(\mathbf t)|\}}\right),
\end{equation*}
and 
\begin{equation*}
\theta_{\mathbf{e}_2}(\mathbf{t}, \Delta)= L_{\mathbf{u}_1}(\mathbf{t}) |\sin(\widetilde \alpha (\mathbf{t}) )\Delta|^{2H_1(\mathbf t)}+L_{\mathbf{u}_2} (\mathbf{t})|\cos(\widetilde \alpha (\mathbf{t}) )\Delta|^{2H_2(\mathbf t)} + O \left(\Delta^{2\underline H(\mathbf t)+\{\zeta\wedge |H_1(\mathbf t)- H_2(\mathbf t)|\}}\right),
\end{equation*}
since $|H_1(\mathbf t)- H_2(\mathbf t)|< 1$.
Taking the ratios between the two quantities above, we obtain 
\begin{multline*}
\frac{\theta_{\mathbf{e}_2}(\mathbf{t}, \Delta)}{\theta_{\mathbf{e}_1}(\mathbf{t}, \Delta)} = |\tan(\widetilde \alpha (\mathbf{t}) )|^{2\underline H(\mathbf t)} \mathbb{I}_{\{H_1(\mathbf t)<H_2(\mathbf t) \}} \\ +  |\cot(\widetilde \alpha (\mathbf{t}) )|^{2\underline H(\mathbf t)} \mathbb{I}_{\{H_1(\mathbf t)>H_2(\mathbf t) \}} + O \left(\Delta^{\{\zeta\wedge |H_1(\mathbf t)- H_2(\mathbf t)|\}}\right).
\end{multline*}
Noting $ |\tan(\widetilde \alpha (\mathbf{t}) )| = |\tan(\widetilde \alpha (\mathbf{t}) +\pi/2)|$ and $ |\cot(\widetilde \alpha (\mathbf{t}) )|=  |\cot(\widetilde \alpha (\mathbf{t}) +\pi/2)|$ completes the proof.
\end{proof}

The remainder term $O\left(\Delta^{\zeta \wedge |H_1(\mathbf{t}) -H_2(\mathbf{t})|}  \right)$ will be discussed in Section \ref{sec:angle-cor}. Proposition \ref{prop:alpha-prop} allows us to find a function of the angle by examining the ratios of mean-squared variations, as in \eqref{eq:mean-squared-variation}, by searching \textit{solely along the canonical basis}. In order to focus on the main idea of directional regularity, we will consider the case where $H$ and $\alpha$ are constant over $\mathcal{T}$.   Under this condition of $H$ and $\alpha$ being constant over $\mathcal{T}$, the function $g$ defined in Proposition \ref{prop:alpha-prop} will be constant over $\mathcal T$ as well.    Hereafter, the notational dependence of $H$, $\alpha$ and  $g$   on $\mathbf{t}$ will be suppressed.   For instance, $g(\alpha(\mathbf t), \mathbf t)$ will simply be denoted  by $g_{\alpha}$. When the directional regularity is constant over $\mathcal{T}$, more stable estimates can be obtained by   averaging over a discrete grid $\widetilde{\mathcal{T}}$.   

The estimation methodology for directional regularity is decomposed into two steps. Firstly, we establish a proxy for the regularity in a given direction. Secondly, this proxy is directly estimated from the data. Formally, for a fixed angle $\beta \in [0, 2\pi)$, we define
\begin{equation}\label{proxyregu}
 H_{\mathbf u(\beta)} (\Delta)= \frac{ \log (\theta _{\mathbf u(\beta)}(2\Delta))-\log (\theta _{\mathbf u(\beta)}(\Delta)) }{2\log(2)},
\end{equation}
where $\mathbf{u}(\beta) = \cos(\beta) \mathbf{e}_1 + \sin(\beta) \mathbf{e}_2$, and
\begin{equation*}
\theta_{\mathbf u(\beta)}(\Delta) = \frac{1}{\#\widetilde{ \mathcal T}} \sum_{\mathbf t\in \widetilde{ \mathcal T}} \theta_{ \mathbf{u}(\beta)  } (\mathbf t, \Delta).
\end{equation*}
  The derivation of \eqref{proxyregu} is given in the Supplementary Material \citep{kassiwangsupp25}. In particular, the proxy avoids the estimation of $L_{\mathbf{u}(\beta)}$ since it is canceled by taking differences of the log terms.   

Whenever we write $H_{\mathbf u(\beta)}(\Delta)$ with an explicit dependence on $\Delta$, or mention the proxy, we are referring to the quantity in \eqref{proxyregu}. We start by providing a bound for this quantity for different angles, which provides us with important insight into the behavior of the directional regularity in the "real world", where $\Delta$ cannot be infinitesimally small. See Section \ref{sec:discussion} for a more detailed discussion. Recall that we are working with $G_{ \mathbf{u}(\beta)  }(\mathbf t , \Delta)= O(\Delta ^{H_{\mathbf u(\beta)}+\zeta })$; see Definition \ref{def:direg-def}. The approximation of the regularity along $\mathbf{u} (\beta)$ in \eqref{proxyregu} serves as a good proxy that converges to the true value of $H_{\mathbf{u} (\beta)}$ as $\Delta$ approaches zero. A second advantage of the proxy in \eqref{proxyregu} is that the map $\beta \mapsto  H_{\mathbf{u}(\beta)}(\Delta)$ is continuous for a fixed $\Delta$,   (see Proposition \ref{prop:continuity_proxy})   , which is not the case for the map $\beta \mapsto H_{\mathbf{u}(\beta)}$, as shown by Lemma \ref{lm}.
\begin{proposition}\label{prop:continuity_proxy}
It holds that
\begin{equation*}
\left|H_{\mathbf u (\beta)} - H_{\mathbf u (\beta)}(\Delta)\right| = O\left(\Delta^{\zeta} \right).
\end{equation*}
 
Suppose that the following continuity condition for $X$ holds true: a positive constant $\mathfrak{a}$ exists such that, for any $\Tt, \mathbf s \in\mathcal T$, 
	\begin{equation}
	\EE\left| 	X^{(j)}\left(\Tt\right)-
	 X^{(j)}\left(\Ss
	 \right)\right|^{2}\leq  \mathfrak{a}  \|\Tt-\Ss\|^{2\underline H }.
	\end{equation}
  
For any pair of angles $(\beta_1, \beta _2) \in [0, 2\pi)^{ 2   }$,  let $L_{\mathbf u(\beta_i)}=(\# \widetilde {\mathcal T})^{-1} \sum_{t\in \widetilde {\mathcal T}} L_{\mathbf u(\beta_i)}(\mathbf t), i = 1, 2$. Then, for any $\Delta$ sufficiently small, it holds that 
\begin{equation}\label{continuity_of_proxy}
\frac{\left|H_{\mathbf u(\beta_1)}(\Delta) - H_{\mathbf u(\beta_2)}(\Delta) \right|}{ |\beta _1 - \beta_2| ^{2\underline H}}\leq \mathfrak a \sum_{k=1,2}\frac{ 2^{\underline H -2H_{\mathbf u(\beta_k)} }}{\log(2)L_{\mathbf u(\beta_k)}} \Delta^{2(\underline H - H_{\mathbf u(\beta_k)})}  \left\{ 1+ O\left(\Delta^\zeta \right)\right\}.  
\end{equation}  
\end{proposition}
The proof of Proposition \ref{prop:continuity_proxy} is given in the Supplementary Material \citep{kassiwangsupp25}. Due to the unique replication nature of functional data, the quantities in \eqref{eq:g-alpha-true} can be easily estimated,   by the sample mean. A natural plug-in estimator for the mean-squared variations is its empirical counterpart, given by
\begin{equation}\label{eq:theta-check}
\check \theta_{\mathbf{e}_i}(\mathbf{t}, \Delta) =  \frac{1}{N}\sum_{j=1}^N \left\{\widetilde X^{(j)}\left(\mathbf{t} - (\Delta/2)\mathbf{e}_i\right) - \widetilde X^{(j)}\left(\mathbf{t} + (\Delta/2)\mathbf{e}_i\right) \right\}^2,
\end{equation}
where $\widetilde X^{(j)}$ denotes an observable approximation of $X^{(j)}$.   For example, $\widetilde X^{(j)}$ can be obtained using interpolation or a kernel smoother applied to the observed data $(Y_m^{(j)}, \mathbf{t}_m), 1 \leq m \leq M_0$. See Remark \ref{rmk:smooth-risk}.    Averaging over   a discrete grid,   we obtain 
\begin{equation}\label{eq:theta-estim}
\widehat \theta_{\mathbf{e}_i}(\Delta) = \frac{1}{ \#  \widetilde{\mathcal{T}}}\sum_{\mathbf{t} \in \widetilde{\mathcal{T}}} \check \theta_{\mathbf{e}_i}(\mathbf{t}, \Delta) - 2\widehat \sigma^2, \qquad i = 1, 2,
\end{equation}
where 
\begin{equation}\label{eq:sigma-estim}
\widehat \sigma^2 = \frac{1}{M_0}\sum_{m=1}^{M_0} \frac{1}{2N}\sum_{j=1}^N \left(Y^{(j)}(\mathbf{t}_{m}) - Y^{(j)}(\mathbf{t}_{m, 1}) \right)^2,
\end{equation}
is an estimator of the noise term in \eqref{eq:data-model}, with $\mathbf{t}_{m,1}$ denoting the closest observed point to $\mathbf{t}_m$. Since the mean-squared variations are associated with the true realizations of the process $X$ and not the observed, noisy version, denoising is important to obtain an estimator with good rates of convergence. 

\begin{remark}\label{rmk:smooth-risk}
In principle, one can choose between a variety of methods to construct the approximations $\widetilde X^{(j)}$. Only a mild moment condition of the form $R_2(M_0) \lesssim M_0^{-\nu}
$, for some $\nu > 0$ is required, where $R_p(M_0) = \sup_{\mathbf{t} \in \mathcal{T}} \mathbb{E}[|\widetilde X_j(\mathbf{t}) - X_j(\mathbf{t})|^p]$. This is satisfied by many non-parametric smoothers, such as kernel smoothers and series expansions. See for instance \citep{fan2016multivariate} and \citep{BELLONI2015}. We adopt nearest-neighbor interpolation, breaking ties by lexicographic order.   For example, for $\mathbf{t} \in \mathcal{T}$, if $\mathbf{t}_m^{(j)}, \mathbf{t}_{m^\prime}^{(j)}$ are equidistant to $\mathbf{t}$, then $Y_{\min\{m, m^{\prime}\}}^{(j)}$ is used for constructing $\widetilde X^{(j)}(\mathbf{t})$ since it is sufficient for our purposes, when coupled with a denoising step as in \eqref{eq:theta-estim}.

Nearest-neighbor interpolation has the advantage of requiring no tuning parameters and is computationally efficient. 
\end{remark}

In order to provide non-asymptotic bounds for our main algorithms, as well as the auxiliary estimates, the following mild assumptions are imposed. Detailed proofs are provided in the Supplementary Material \citep{kassiwangsupp25}.

\begin{enumerate}[label=H\arabic*]
	\item\label{ass_D}  Let $X$ be an anisotropic process with two regularities $(H_1,H_2)$, where $X^{(j)}$, $1\leq j \leq N$,  are independent realizations of $X$.
	\item  \label{ass_D1} The error terms  $\varepsilon^{(j)}_m$ in equation \eqref{eq:data-model} are i.i.d, and independent from the process $X$.	

	\item\label{ass_H1}  Three positive constants $\mathfrak{a}$, $\mathfrak{A}$ and $r$ exist such that, for any $\Tt\in\mathcal T$,  
	$$
	\EE\left| 	X^{(j)}\left(\Tt\right)-
	 X^{(j)}\left(\Ss
	 \right)\right|^{2p}\leq  \frac{p!}{2}\mathfrak{a} \mathfrak{A}^{p-2} \|\Tt-\Ss\|^{2p\underline H },
	\qquad \forall \Ss\in B(\Tt; r) ,\; \forall p\geq 1.
	$$
		\item\label{ass_noise} A constant $\mathfrak{G}$ exists such that 
	\begin{equation*}
	\mathbb{E}(\{\varepsilon_1^{(1)} \}^{2p} ) \leq \frac{p!}{2}\mathfrak{G}^{p-2} \sigma^2, \qquad \forall p \geq 1.
	\end{equation*}
\end{enumerate}

\begin{remark}
Assumption (\ref{ass_H1}) states that $X$ has sub-Gaussian increments in a local sense, for every point $\mathbf{t} \in \mathcal{T}$.   It is a stronger version than the one required in Proposition \ref{prop:continuity_proxy}.     It is satisfied by the processes considered in the simulations, as well as those presented in Section \ref{sec:examples}. In fact, under Definition \ref{def:direg-def}, Assumption  (\ref{ass_H1}) is equivalent to 
$$
\EE\left| 	X^{(j)}\left(\Tt\right)-
	 X^{(j)}\left(\Ss
	 \right)\right|^{2p}\leq  \frac{p!}{2}\mathfrak{a} \mathfrak{A}^{p-2} \EE\left[\left| 	X^{(j)}\left(\Tt\right)-
	 X^{(j)}\left(\Ss
	 \right)\right|^{2}\right]^p,
$$
an assumption which is more familiar and widely used in the literature. 
\end{remark}

A recent line of work in regularity estimation within the context of FDA has emerged, which we can exploit for our purposes. See for example \citep{Golovkine2022,Golovkine2021,wang2023adaptive,maissoro2024,kassi2023,hsing2020}.
An estimate of the regularity $H$ along an arbitrary direction $\mathbf{u}(\beta)$ can be obtained by estimating the proxy \eqref{proxyregu} as follows:
\begin{equation}\label{eq:H_estim_dir}
\widehat H_{\mathbf{u}(\beta)}(\Delta) = \begin{cases}
\frac{\log(\widehat \theta_{\mathbf{u}(\beta)}( 2\Delta)) - \log(\widehat \theta_{\mathbf{u}(\beta)}( \Delta))}{2\log(2)} \qquad \text{if} \qquad \widehat{\theta}_{\mathbf{u}(\beta)} ( 2\Delta) \geq \widehat{\theta}_{\mathbf{u}(\beta)} ( \Delta) > 0, \\
1 \qquad \text{otherwise}.
\end{cases}
\end{equation}
Proposition \ref{prop:conc:uni:beta} is a critical ingredient to derive rates of convergence associated with the identification process in Algorithm \ref{algo:alpha-hat-algo}, as it makes use of the estimated angles, introducing dependence.

\begin{proposition}\label{prop:conc:uni:beta}
Assume that (\ref{ass_D}), (\ref{ass_D1}), (\ref{ass_H1}) and (\ref{ass_noise}) hold true. Recall that $\underline H = H_1 \wedge H_2$ and $\overline H = H_1 \vee H_2$. For any $\eta$ which satisfies
\begin{equation}\label{eq:eta-bias-unif}
\eta\geq   \frac{4\mathfrak a}{\underline L \Delta^{2\overline H}}\times \left\{  \left(\sqrt{2}\right)^{-\underline H} \left\{ \left(\sqrt{2}\right)^{2-\underline H} M_0^{-\frac{1}{2}\underline H} +2 \Delta^{\underline H}\right\}M_0^{-\frac{1}{2}\underline H} + M_0^{-\underline H}\right\},
\end{equation}
we have for any $r\in [0, \underline H)$,
\begin{align*}
\PP\left( \sup_{\beta \in [0, 2\pi)} |\widehat H_{\mathbf u(\beta)}(\Delta) -  H_{\mathbf u(\beta)}(\Delta)| \geq 2\eta\right) \leq 4\exp \left (\frac{-\eta^2 N\Delta^{4\underline H} }{8\mathfrak b + 4\mathfrak B\eta\Delta^{2\underline H}}\right)
+ 4 \exp\left(\frac{-\eta^2 N\Delta^{4 \underline H} }{64\mathfrak a_1 + 8\mathfrak A_1 \eta \Delta^{2\underline H}} \right),
\end{align*}
where 
$$
\mathfrak A_1 = \frac{4\mathfrak A}{(2M_0)^{\underline H}} \vee 16\mathfrak G, \qquad \mathfrak a_1 = \frac{\mathfrak a}{ (2M_0)^{2\underline H}} + 2^4\sigma^2, \qquad \mathfrak b= \mathfrak B^2 \left( \frac{\mathfrak a}{2 \mathfrak A^2}\right)^{\frac{\underline H-r}{1 -\underline H +r}},
$$
and
\begin{equation*}
 \mathfrak B= \left(\frac{\sqrt 22^{\underline H} \mathfrak A}{\log(2)} \right)^2 (\underline H-r)^{-3}  \left\{\sqrt2 M_0^{-1/2} + \Delta\right\}^{2r}.
 \end{equation*}
\end{proposition}
The proof of Proposition \ref{prop:conc:uni:beta} is given in the Supplementary Material \citep{kassiwangsupp25}. The condition on $\eta$ in \eqref{eq:eta-bias-unif} represents the bias term that arises from performing denoised interpolation in order to compute the mean-squared variations. This bias term is an intrinsic feature of the "discretely observed" setup, where the surfaces are observed only at a finite number of discrete points.  
The first exponential term in the bound corresponds to the stochastic error arising from estimating the mean-squared variations using the sample mean, whereas the second exponential term accounts for the stochastic error in estimating the variance of the measurement errors $\varepsilon_m^{(j)}$.
  The constants $\mathfrak A_1$ and $\mathfrak a_1$ are associated with estimating the variance of the noise, whereas $\mathfrak B$ and $\mathfrak b$ arise from the estimation of the mean-squared variations. It is worth noticing that $ \mathfrak B$ is tending to zero as $M_0 \rightarrow \infty$ and $\Delta \rightarrow 0$, implying that the second exponential term (the noise contribution) dominates the bound. However, in the noiseless case ($\sigma =0$), assumption (\ref{ass_noise}) is satisfied with $\mathfrak G =0$. This implies on the one hand that $ \mathfrak a_1 \asymp M_0^{-2\underline H}$ and $\mathfrak A_1 \asymp M_0^{- \underline H}$, while, on the other hand, $ \mathfrak b \asymp M_0^{-2r}+ \Delta^{4r}$ and $\mathfrak B \asymp M_0^{- r}+ \Delta^{2r}$ with $r<\underline H$. Consequently, the first exponential term dominates the bound in the absence of noise.

The following corollary is a direct consequence of Proposition \ref{prop:conc:uni:beta}, with an appropriate choice of $\Delta$. 

\begin{corollary}\label{para:setting}
Suppose that the assumptions of Proposition \ref{prop:conc:uni:beta} hold true, and set  
\begin{equation}\label{deltachoice}
\Delta = \exp\left(-\log^{\xi}(M_0) \right), \qquad \text{for } 0<\xi<1.
\end{equation}
Moreover, we assume that a non-negative constant $\mathfrak e >0$ (independent of $M_0$ and $N$) exists such that 
\begin{equation}\label{M_0-N:com}
\mathfrak e^{-1} \leq \frac{\log(M_0)}{\log(N)} \leq \mathfrak e.
\end{equation}
Then
\begin{equation}\label{rate:dh}
\sup_{\beta \in [0, 2\pi)} |\widehat H_{\mathbf u(\beta)}(\Delta) -  H_{\mathbf u(\beta)}(\Delta)|= O_{\PP}\left(\frac{ \exp\left((2\overline H - \underline H) \log^{\xi}(M_0)\right)}{\sqrt{ M_0 }^{\underline H}} \vee \frac{ \exp\left(2 \underline H \log^{\xi}(M_0)\right)}{\sqrt N}  \right).
\end{equation}
\end{corollary}
The condition \eqref{M_0-N:com} requires that $M_0$ falls within two powers of $N$, or equivalently, that the converse holds true. The choice of $\Delta $ in \eqref{deltachoice} is done in such a way that $\log^{-b}(M_0) \gg \Delta \gg M_0^{-b}$ for any fixed $b>0$. This choice can also be found in \citep{Golovkine2022} with $\xi = 1/3$. The rate terms in \eqref{rate:dh} converge to $0$ since the choice of $\Delta $ is  larger than any negative power of $M_0$, and keeping \eqref{M_0-N:com} in mind.

The minimum regularity can be estimated as 
\begin{equation}\label{eq:H-min-estim}
\widehat{\underline{H}}(\Delta) = 
\begin{cases}
\min_{i=1, 2} \frac{\log(\widehat \theta_{\mathbf{e}_i}(2\Delta)) - \log(\widehat \theta_{\mathbf{e}_i}(\Delta))}{2\log(2)} \qquad \text{if} \qquad \widehat{\theta}_{\mathbf{e}_i} ( 2\Delta), \widehat{\theta}_{\mathbf{e}_i} (\Delta) > 0, \\
1 \qquad \text{otherwise}.
\end{cases}
\end{equation}
The regularity estimator in \eqref{eq:H-min-estim} is a slight adaptation of \citep{kassi2023} by taking the minimum over the index of the basis vectors. This is motivated by Lemma \ref{lm}, which says that at least one of the two vectors gives us the smallest regularity for the process $X$. Collecting \eqref{eq:theta-estim}, \eqref{eq:sigma-estim} and \eqref{eq:H-min-estim}, a plug-in estimator of \eqref{eq:g-alpha-true} is given by 
\begin{equation}\label{eq:g-alpha-estim}
 \widehat{g_{\alpha}}( \Delta)   = 
\left(\frac{\widehat \theta_{\mathbf{e}_2}(\Delta)\mathbb{I}_{\widehat \theta_{\mathbf{e}_2}(\Delta) > 0} + \mathbb{I}_{\widehat \theta_{\mathbf{e}_2}(\Delta) \leq 0}}{\widehat \theta_{\mathbf{e}_1}(\Delta)\mathbb{I}_{\widehat \theta_{\mathbf{e}_1}(\Delta) > 0} + \mathbb{I}_{\widehat \theta_{\mathbf{e}_1}(\Delta) \leq 0}} \right)^{1/\{2\widehat{\underline{H}}(\Delta)\}},
\end{equation}
  where $\widehat{g_{\alpha}}( \Delta)$ represents an estimator of a transformation of $\alpha$, since our methodology does not yield a direct estimator for $\alpha$ itself.   By construction, the only relevant tuning parameter in \eqref{eq:g-alpha-estim} is the spacing $\Delta$. We address this issue in Section \ref{sec:alpha-iden}, immediately after Algorithm \ref{algo:alpha-hat-algo} is presented. Proposition \ref{g:concentration} provides a concentration inequality for the estimates in \eqref{eq:g-alpha-estim}.
\begin{proposition}\label{g:concentration}
Under the assumptions of Corollary \ref{para:setting}, three positive constants $C_1$, $C_2$ and $\mathfrak u$ exist such that for any   $\eta$    satisfying
\begin{equation}\label{consitionep}
\frac{|\log(\Delta)|}{\mathfrak u \Delta^{2\underline H}}\geq \eta  \geq\mathfrak u\max\{ \Delta^{-3\underline H } |\log(\Delta)| M_0^{ - \frac{1}{2} \underline H} , \Delta ^{\zeta\wedge|H_1-H_2|}\},
\end{equation}
 we obtain
\begin{equation}\label{probaghat}
\PP \left(| \widehat{g_{\alpha}}( \Delta) - g_{\alpha} |\geq  \eta   \right ) \leq C_1 \exp\left(- C_2 \eta  ^2N\frac{\Delta^{8\underline H}}{\log^2(\Delta)} \right),
\end{equation}
with $g$ defined as in Proposition \ref{prop:alpha-prop}.
\end{proposition}
The proof is given in the Supplementary Material \citep{kassiwangsupp25}. The condition on  $\eta$   in \eqref{consitionep} represents the remainder term in Proposition \ref{prop:alpha-prop}. Combined with the condition \eqref{eq:eta-bias-unif} on $\eta$ in Proposition \ref{prop:conc:uni:beta}, such a condition on  $\eta$   can be satisfied by choosing $\Delta$ as in \eqref{deltachoice}, provided that $M_0$ is sufficiently large. The probability bound in \eqref{probaghat} converges to zero as long as we choose $\Delta$ as in \eqref{deltachoice}, and the condition \eqref{M_0-N:com} is satisfied.

\subsection{Identification issues}\label{sec:alpha-iden}
The estimating equation in \eqref{eq:g-alpha-true} reveals two identification problems. The first issue is related to the function $g$, which can either be the tangent ($\tan$) or cotangent ($\cot$) function depending on the ordinal nature of $H_1$ and $H_2$. This reflects the phenomenon that the dominating term arising from the ratio of mean-squared variations differs, depending on whether $H_1 > H_2$. This ordinality is unknown a priori.  

The second issue is associated with the absolute value on the LHS of \eqref{eq:g-alpha-true}. By using the estimator in \eqref{eq:g-alpha-estim}, one obtains either $ g_{\alpha}  $ or $ g_{\pi - \alpha} $. Similarly, the sign is unknown a priori to the practitioner. Fortunately, an identification procedure can be constructed to resolve these two issues. 

Let $\mathbf{u}$ be a unit vector represented in the canonical basis as $\mathbf{u}(\beta) = \cos(\beta)\mathbf{e}_1 + \sin(\beta)\mathbf{e}_2$. By construction, the regularity $H_{\mathbf{u}(\alpha)}$ is maximal when built using the angle $\alpha$ in \eqref{eq:g-alpha-true}. The true value of $\alpha$ can thus be identified as
\begin{equation}
\alpha = {\arg\max}_{\beta \in \left\{\gamma, \pi - \gamma, \pi/2 - \gamma, \pi/2 + \gamma \right\}} H_{\mathbf{u(\beta)}},
\end{equation}
where 
\begin{equation}
\gamma \approx \gamma(\Delta) = \text{arccot}\left(\frac{\theta_{\mathbf{e}_2}(\Delta)}{\theta_{\mathbf{e}_1}( \Delta)}\right)^{\frac{1}{2 \underline H( \Delta)}}.
\end{equation}
The angles $\gamma$ are computed by applying the relevant inverse maps to \eqref{eq:g-alpha-estim}. In order to obtain more robust estimates $\widehat H_{\mathbf{u}(\beta)}$ with respect to the spacing $\Delta$, we propose to compute $\widehat H_{\mathbf{u}(\beta)}$ over a grid of points $\boldsymbol{\Delta} = (\Delta_1, \dots, \Delta_{K_0})^{\top}$.
The angle which maximizes the sum over $\boldsymbol{\Delta}$ is then selected, yielding
\begin{equation}\label{eq:alpha-iden-hat}
\widehat \alpha = \arg\max_{\beta \in \left\{\widehat \gamma, \pi - \widehat \gamma, \pi/2 - \widehat \gamma, \pi/2 + \widehat \gamma \right\}} \sum_{k=1}^{K_0} \widehat H_{\mathbf{u}(\beta)}(\Delta_k).
\end{equation}
As seen from Theorem \ref{alpha:concen}, the spacing $\Delta$ used for identification should be chosen such that it is slower in rate compared to the one used for estimation, providing additional justification for \eqref{eq:alpha-iden-hat}. The estimation and identification procedure are summarized in  Algorithm \ref{algo:alpha-hat-algo}. Theorem  \ref{alpha:concen} provides a concentration inequality for the identification procedure.

\begin{theorem}\label{alpha:concen}
Suppose that the assumptions of Proposition \ref{prop:conc:uni:beta} hold true. Moreover assume that for any $k\in \{1, \dots, K_0 \}$, we have 
\begin{equation}\label{conditiondeltak}
\Delta=o\left( \Delta_k^{\frac{\overline H- \underline H}{ \left(  \zeta \wedge\{\overline H - \underline H\}  \right)  \underline H }}\right), \qquad \text{as } \Delta \rightarrow 0.
\end{equation} 
Then for any $ \eta  $ such that $\Delta^{\zeta\wedge\{\overline H - \underline H\}} \ll  \eta $, we have for $M_0$ sufficiently large,
\begin{multline}\label{alphaconv}
\mathbb P \left(|\widehat \alpha - \alpha|\geq2  \eta  \right)\leq  \PP\left( \sup_{\beta \in [0, 2\pi)} |\widehat H_{\mathbf u(\beta)}(\Delta) -  H_{\mathbf u(\beta)}(\Delta)| \geq \frac{\overline H - \underline H}{8K_0}\right) \\ + \PP \left(| \widehat{g_{\alpha}}( \Delta) - g_{\alpha} |\geq  \eta  \right )
+ \PP \left(| \widehat{g_{\alpha}}( \Delta)  - g_{\alpha} |^{2\underline H}\geq \frac{ \underline L(\overline H - \underline H)}{ 4\mathfrak a 2^{3- \underline H}K_0} \right).
\end{multline} 
\end{theorem}

The proof is given in the Supplementary Material \citep{kassiwangsupp25}. The convergence of the RHS in \eqref{alphaconv} can be guaranteed following the discussion of Propositions \ref{prop:conc:uni:beta}, \ref{g:concentration} and Corollary \ref{para:setting}. As seen in \eqref{conditiondeltak}, in order to identify the right angle which gives the maximal regularity, we need to choose the values in the grid $\boldsymbol \Delta$, introduced in \eqref{eq:alpha-iden-hat}, to have a slower rate of decrease  than the $\Delta$ used to estimate the function $g_{  \alpha  }$ in \eqref{eq:g-alpha-estim}.   In practice, for $k\in \{1, \dots, K_0\}$, one can take $\Delta_k = \Delta^{\tau_k}$, with $\tau_k \{\overline H - \underline H \}\{\zeta \wedge (\overline H - \underline H)\underline H \}^{-1}< 1$. Thus, at the very least, $\tau_k < 1$ since $(\zeta \wedge\{\overline H - \underline H\})\{\overline H- \underline H\}^{-1} \underline H < 1$. Since the term multiplying $\tau_k$ is unknown in practice, choosing $\tau_k < 1$ is a simple and sensible choice. For example, we can choose $\tau_k= 0.5 + 0.49/k$ in practice.

\begin{algorithm}
\caption{Estimation and identification of $\alpha$}
\label{algo:alpha-hat-algo}
\begin{algorithmic}[1]
\Require Data $(Y^{(j)}(\mathbf{t}_m), \mathbf{t}_m)$, Grid $\widetilde{\mathcal{T}} = \left\{\mathbf{t}_1, \dots, \mathbf{t}_p \right\}$,   Spacing $\Delta$, Grid of spacings $\boldsymbol{\Delta}$;   

Initialize $\widehat H_{\mathbf{u}(\beta)} \gets \emptyset$;

\State Compute $ \widehat{g_{\alpha}}( \Delta) $ according to \eqref{eq:g-alpha-estim};

\State $\widehat \alpha^{tan} \gets \arctan\left( \widehat{g_{\alpha}}( \Delta) \right)$;

\State $\widehat \alpha^{cot} \gets \text{arccot} \left( \widehat{g_{\alpha}}( \Delta)  \right)$;

\State $\mathbf{u}(\beta) \gets (\cos(\beta), \sin( \beta))^{\top}$;\Comment $\forall \beta \in \{\widehat \alpha^{tan}, \widehat \alpha^{cot}, \pi - \widehat \alpha^{tan}, \pi - \widehat \alpha^{cot}\}$;

\For{{$\Delta_k$ \texttt{in} $\boldsymbol{\Delta}$}}

\State Compute $\widehat H_{\mathbf{u}(\beta)}(\Delta_k)$ according to \eqref{eq:H_estim_dir};

\State $\widehat H_{\mathbf{u}(\beta)} \gets  \widehat H_{\mathbf{u}(\beta)} \bigcup \widehat H_{\mathbf{u}(\beta)}(\Delta_k)$;

\EndFor

Compute $\widehat \alpha$ according to \eqref{eq:alpha-iden-hat};

\State \Return $\widehat \alpha$
\end{algorithmic}
\end{algorithm}

In the estimation of $ \widehat{g_{\alpha}}( \Delta)  $, it is sufficient to choose a fixed spacing $\Delta$. In order to ensure that the nearest-neighbor of $\mathbf t+( \Delta/2) \mathbf{e}_i$ is distinct to the nearest-neighbor of $\mathbf t -( \Delta/2) \mathbf{e}_i$ when computing $\theta_{\mathbf{e}_i}$ in $ \widehat{g_{\alpha}}( \Delta)  $, $\Delta$ should be chosen at least as large as $(2 M_0)^{-1/2}$. However we need $\Delta$ to be strictly larger than the minimum size required to capture one point in order to obtain asymptotic concentration. In the non-asymptotic setting, we propose a conservative choice of $\Delta = M_0^{-1/4}$. Simulation results in Section \ref{sec:numerics} confirm that this choice works well in practice.

\subsection{Correction for the remainder term}\label{sec:angle-cor}
\begin{figure}[h]
\centering
\includegraphics[scale=0.5]{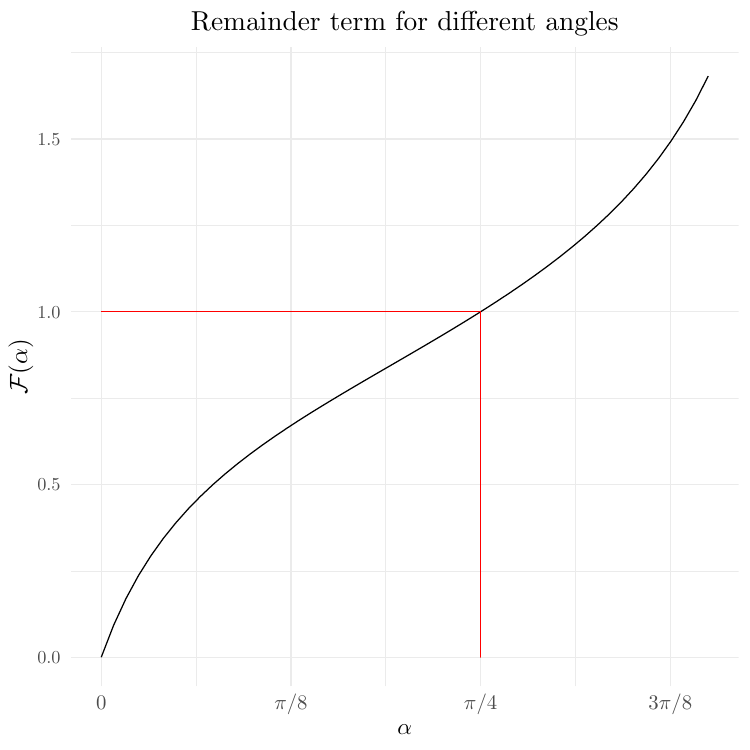}
\caption{Plot showing the remainder term for different angles $\alpha$. $\mathcal F(\alpha) \rightarrow \infty$ as $\alpha \rightarrow \pi/2$. Going closer to the boundary $0$ and $\pi/2$ can dramatically affect the estimation of the true angles.}
\label{fig:f-remain_plot}
\end{figure}

Let $\widetilde {\mathcal{F}}(\alpha) =\widetilde{  \mathcal{F}}(\alpha, \zeta) = O\left(\Delta^{\zeta \wedge |H_1 -H_2|} \right)$ denote the remainder term in \eqref{eq:g-alpha-true}. It turns out that $\widetilde {\mathcal{F}}$ can become big enough to affect the estimation of $g(\alpha)$ through the constants, which is more pronounced for certain angles $\alpha$. Let $L_{\overline{\mathbf{u}}}$  be the local  Hölder constant in the direction of the maximizing  regularity $\overline H$, and let $L_{\underline{\mathbf{u}}}$ be the local Hölder constant in the orthogonal direction of $\overline {\mathbf u}$. The explicit form of the remainder term is given by  
\begin{equation}\label{eq:g-remainder}
\widetilde {\mathcal{F}}(\alpha)  = 
\left(\frac{1 + \mathcal{F}_{num}}{1 + \mathcal{F}_{denom}}  \right)^{\frac{1}{2\underline H}} + O\left(  \Delta^{\zeta \wedge1}\right)=: \mathcal{F}(\alpha) + O\left(  \Delta^{\zeta\wedge 1}\right)  ,
\end{equation}   
where 
\begin{equation*}
\mathcal{F}_{num} = \frac{1}{L_{\underline{\mathbf{u}}}(\Delta)}
\frac{L_{\overline{\mathbf{u}}}(\Delta)\left|\cos(\alpha)\mathbb{I}_{\{H_1 < H_2\}} + \sin(\alpha)\mathbb{I}_{\{H_1 > H_2\}}\right|^{2\overline H}\Delta^{2\overline H} + R(\Delta)}{\left|\sin(\alpha)\mathbb{I}_{\{H_1 < H_2 \}} + \cos(\alpha)\mathbb{I}_{\{H_1 > H_2 \}}  \right|^{2\underline H}\Delta^{2\underline H}},
\end{equation*}
\begin{equation*}
\mathcal{F}_{denom} = \frac{1}{L_{\underline{\mathbf{u}}}(\Delta)} \frac{L_{\overline{\mathbf{u}}}(\Delta)\left|\sin(\alpha)\mathbb{I}_{\{H_1 < H_2\}} + \cos(\alpha)\mathbb{I}_{\{H_1 > H_2\}}\right|^{2\overline H}\Delta^{2\overline H }+R(\Delta)}{\left|\cos(\alpha)\mathbb{I}_{\{H_1 < H_2 \}} + \sin(\alpha)\mathbb{I}_{\{H_1 > H_2 \}}  \right|^{2\underline H}\Delta^{2 \underline H}},
\end{equation*}
and 
\begin{multline}\label{croisprod}
R(\Delta)= \mathbb E\left[  \left\{X\left(\mathbf t - \frac{a_1 \Delta}{2} \mathbf{u}_1-\frac{a_2 \Delta}{2} \mathbf{u}_2 \right) -X\left(\mathbf t + \frac{a_1 \Delta}{2} \mathbf{u}_1-\frac{a_2 \Delta}{2} \mathbf{u}_2 \right)  \right\} \right.\\
\left. \times\left\{ X\left(\mathbf t + \frac{a_1 \Delta}{2} \mathbf{u}_1-\frac{a_2 \Delta}{2} \mathbf{u}_2 \right) -X\left(\mathbf t + \frac{a_1 \Delta}{2} \mathbf{u}_1+\frac{a_2 \Delta}{2} \mathbf{u}_2 \right) \right\}\right].
\end{multline}
It is easy to see from \eqref{eq:g-remainder} that $\mathcal{F}(\pi/4) = 1$, regardless of the ordinal nature of $H_1$ and $H_2$. This implies that when $\alpha = \pi/4$, the estimating equations in \eqref{eq:g-alpha-true} are  most accurate. However as $\alpha$ tends to 0 or $\pi/2$, the estimating equations can be dramatically affected by this remainder term. A plot of $\mathcal{F}(\alpha)$ can be seen in Figure \ref{fig:f-remain_plot}. 

The remainder term $R$ is intrinsic to the process $X$. It can be bounded by Cauchy-Schwarz inequality to obtain $R= O\left(\Delta^{ |H_1 -H_2|} \right)$. However, this is a pessimistic approach, since $R$ can be equal to zero, for processes such as \eqref{sum:brows}. In order to adjust for this possibly non-negligible remainder term, the terms $R$ and $\mathcal{F}(\alpha)$ can be estimated, provided that $2(\overline H - \underline H) < \zeta\wedge 1$. Since $R$ only involves expectations of the process, it can be estimated by \eqref{eq:theta-estim}. We now focus on the estimation of $\mathcal{F}$.

Let $\theta_{\overline{\mathbf{u}}}$ and $ \theta_{\underline{\mathbf{u}}}$ be the mean-squared variations such that $H_{\overline{\mathbf{u}}} = \overline{H}$, and $H_{\underline{\mathbf{u}}} = \underline{H}$ respectively. Since
\begin{equation*}
 \frac{L_{\overline{\mathbf{u}}}(\Delta)}{L_{\underline{\mathbf{u}}}(\Delta)}\Delta^{\overline H -\underline H} \approx  \frac{\theta_{\overline{\mathbf{u}}}(\Delta)}{\theta_{\underline{\mathbf{u}}}(\Delta)},
\end{equation*}
we have 
\begin{equation*}
\mathcal{F}_{num} \approx \frac{\theta_{\overline{\mathbf{u}}}(\Delta)}{\theta_{\underline{\mathbf{u}}}(\Delta)} \frac{\left|\cos(\alpha)\mathbb{I}_{\{H_1 < H_2\}} + \sin(\alpha)\mathbb{I}_{\{H_1 > H_2\}}\right|^{2\overline H}}{\left|\sin(\alpha)\mathbb{I}_{\{H_1 < H_2 \}} + \cos(\alpha)\mathbb{I}_{\{H_1 > H_2 \}}  \right|^{2\underline H}} + \frac{R(\Delta)}{ \theta_{\underline{\mathbf u}}(\Delta)},
\end{equation*} 
and 
\begin{equation}\label{eq:F-hat-approx}
\mathcal{F}_{denom} \approx \frac{\theta_{\overline{\mathbf{u}}}(\Delta)}{\theta_{\underline{\mathbf{u}}}(\Delta)} \frac{\left|\sin(\alpha)\mathbb{I}_{\{H_1 < H_2\}} + \cos(\alpha)\mathbb{I}_{\{H_1 > H_2\}}\right|^{2\overline H}}{\left|\cos(\alpha)\mathbb{I}_{\{H_1 < H_2 \}} + \sin(\alpha)\mathbb{I}_{\{H_1 > H_2 \}}  \right|^{2\underline H}}+ \frac{R(\Delta)}{ \theta_{\underline{\mathbf u}}(\Delta)},
\end{equation}
in the sense that the ratios of the LHS and RHS in \eqref{eq:F-hat-approx} goes to 1 in the limit as $\Delta \rightarrow 0$. $\widehat{\mathcal{F}}(\alpha)$ can be computed by replacing $\theta_{\underline{\mathbf{u}}}$, $\theta_{\overline{\mathbf{u}}}$, and $\alpha$ with their estimates obtained by Algorithm \ref{algo:alpha-hat-algo}. An adjusted estimate of $\alpha$ will then be given by
\begin{equation}\label{eq:g-hat-hat}
  \widetilde{g_{\alpha}}( \Delta)   =  \widehat{g_{\alpha}}( \Delta)   / \mathcal{F}(\widehat \alpha),
\end{equation}
where $ \widehat{g_{\alpha}}( \Delta)  $ is computed by \eqref{eq:g-alpha-estim}. The simulation results in Section \ref{sec:numerics} suggest that using the adjusted estimates in \eqref{eq:g-hat-hat} yields significantly better results when $\alpha$ gets further away from $\pi/4$. In order to avoid introducing greater dependence between quantities, we suggest to only compute the adjusted estimates once. Moreover, we suggest to set $R = 0$ in practice to decrease the computational load. This seems to produce good results , even when $R\ne 0$, as seen from the simulation results in the Supplementary Material \citep{kassiwangsupp25}. Likewise, the final estimate $\check{\alpha}$  can be obtained by applying the appropriate inverse function (either $\arctan$ or arccot) obtained by the identification process when computing the initial estimates $\widehat \alpha$.

\section{  Simulations and real data illustration}\label{sec:numerics}
 
In this section, a simple, novel simulator for a class of bivariate anisotropic processes is described. The simulator, along with the other algorithms, are implemented in the \textbf{R} package \texttt{direg}\footnote{Freely accessible at \href{https://github.com/sunnywang93/direg}{https://github.com/sunnywang93/direg}.}. This is followed by numerical experiments and an illustration on a real data set.

\subsection{Anisotropic simulator}\label{sec:simulator}
Let $B_1$, $B_2$ be two processes with regularities $H_1$ and $H_2$ respectively. Suppose without loss of generality that $H_1 \neq H_2$. In the following, our exposition will assume that $B_1$ and $B_2$ are fractional brownian motions (fBms). In principle, our simulation approach can be generalized to other Gaussian processes with stationary increments.

Many methods are available to simulate these types of processes, and we do not attempt to provide an exhaustive list. For some examples, see \citep{woodchan94, stein2002, coeurjolly2018}. A survey can be found in \citep{dieker2004}. In particular, the circulant embedding method described in \citep{woodchan94} is attractive due to its speed, and can be adapted to simulate processes such as the fBm with stationary increments.

The anisotropic component of the simulator will be our main focus, since there is already a wide array of methods available for simulating the individual processes. Let $\mathbf{u}_1$ and $\mathbf{u}_2$ be unit vectors with associated regularities $H_1$ and $H_2$. The vectors can be represented in the canonical basis as $\mathbf{u}_1 = \cos(\alpha)\mathbf{e}_1 + \sin(\alpha)\mathbf{e}_2$ and $\mathbf{u}_2 = -\sin(\alpha)\mathbf{e}_1 + \cos(\alpha)\mathbf{e}_2$. The main difficulty of simulating processes on an equally spaced grid along $\mathbf{u}_1$, $\mathbf{u}_2$ is the existence of negative values when $\alpha \in [0, 3\pi/2]$, while the fBm has a domain in $\mathbb{R}_+$. 

This problem can be resolved by exploiting the stationary increments property of the fBm, given by $B(t) - B(s) \stackrel{d}{=}B(t-s)$. Using the reference point $t = 0$, we obtain 
\begin{equation}\label{eq:stat-incre}
-B(s) \stackrel{d}{=} B(-s).
\end{equation}
An anisotropic fBm can thus be simulated on any $\alpha \in [0, \pi]$ by first simulating a fBm on an equally spaced grid in $[0, |\cos(\alpha)| + \sin(\alpha)]$, before transforming the negative part using \eqref{eq:stat-incre}. A bivariate process can finally be constructed by applying a function $f$ to the individual processes, for example $f(B_1, B_2) = B_1 + B_2$. Our simulator is similar in spirit to the Turning bands approach; see for example \citep{matheron1973} and \citep{richards2015}. Pseudocode for our simulator can be found in Algorithm   SM.1    of the Supplementary Material \citep{kassiwangsupp25} .

\subsection{Parameter settings and error measures}\label{sec:para}
Observations $(Y^{(i)}(\mathbf{t}_m), \mathbf{t}_m), 1 \leq i \leq N, 1 \leq m \leq M_0$ were simulated using our algorithm for the sum of two fBms. Other processes (e.g., product of two fBms) can be found in the Supplementary Material \citep{kassiwangsupp25}. 

A total of 60 different parameter configurations were explored, consisting of all possible combinations of the following parameter sets: number of curves $N \in \{100, 150\}$, number of points along each curve $M_0 \in \{51^2, 101^2\}$, noise level $\sigma \in \{0.1, 0.5, 1\}$, and angles $\alpha \in \{\pi/30, \pi/5, \pi/4, \pi/3, \pi/2 - \pi/30\}$. In both processes, the regularities were fixed to be $H_1 = 0.8$ and $H_2 = 0.5$.
In line with our discussion in Section \ref{sec:alpha-iden}, the parameter $\Delta$ was set to be 
$\Delta = M_0^{-1/4}$. The grid of spacings $\boldsymbol{\Delta}$ involved in the identification process was set to $\boldsymbol{\Delta} = \{M_0^{-1/4}, \Delta_1, \dots, \Delta_{K_0 - 1},  0.4 \}$, an evenly spaced grid consisting of $K_0 = 15$ points. The absolute error is used as a risk measure for each experiment, given by

\begin{equation}\label{eq:risk-alpha}
\mathcal R_{\alpha} = |\check \alpha - \alpha|, 
\end{equation}
where $  \check{\alpha}$ is the adjusted estimate in \eqref{eq:g-hat-hat}. Parameter settings for anisotropic detection is in Table \ref{table:ani-detect}. 

\subsection{Empirical results}\label{sec:emp-results}
\begin{figure}[!ht]
\centering
\begin{minipage}[b]{0.7\linewidth}
    \includegraphics[width=.45\linewidth]{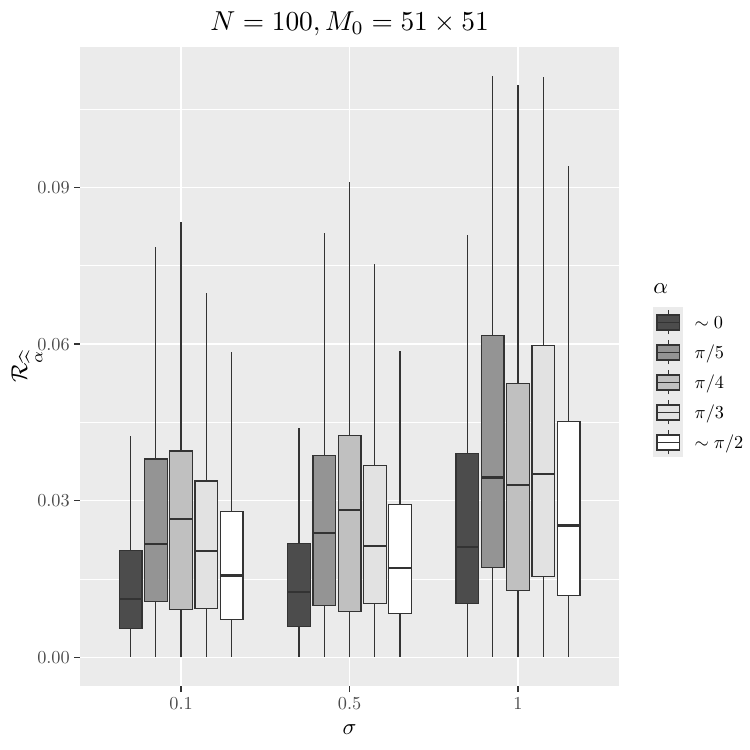} 
    \includegraphics[width = .45\linewidth]{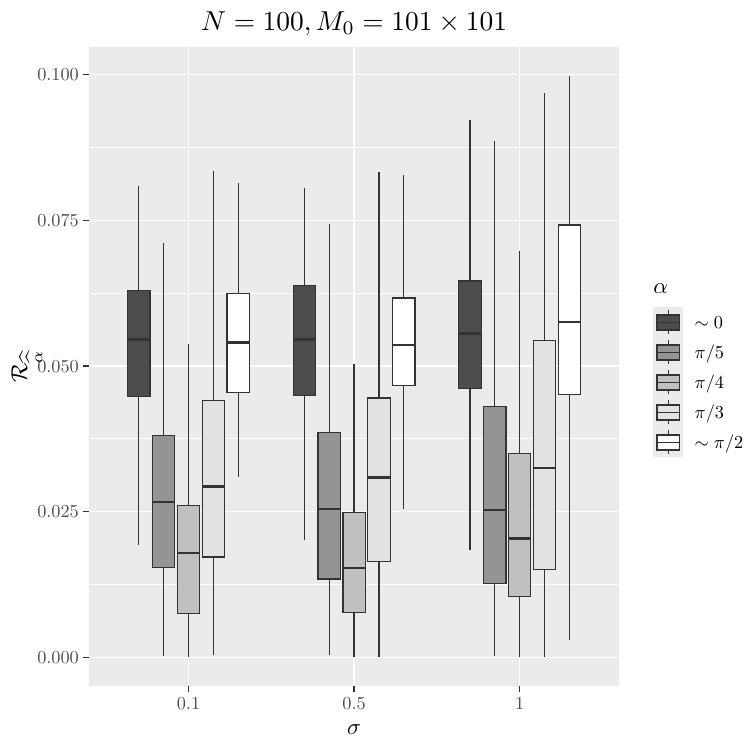}
\includegraphics[width=.45\linewidth]{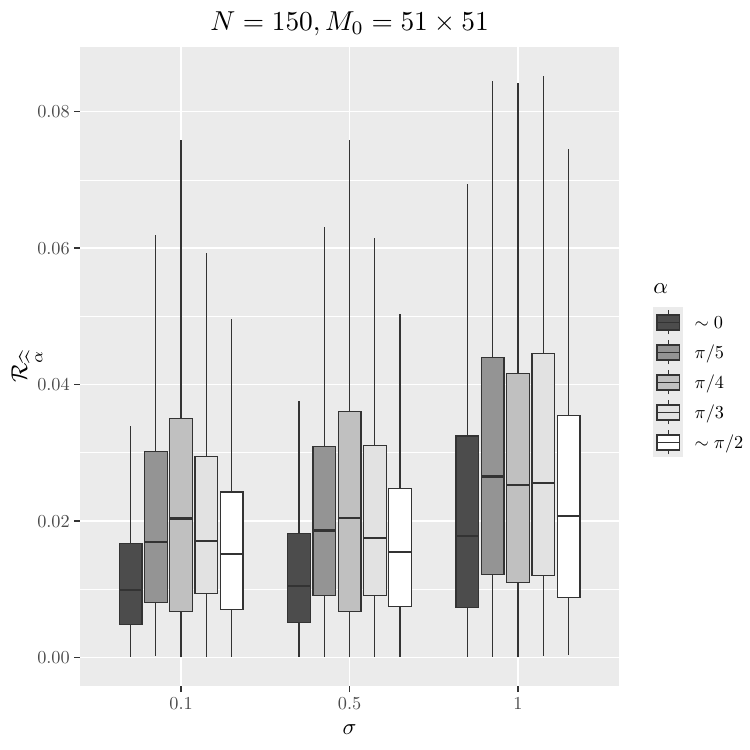}
\hspace{0.9 cm}
\includegraphics[width=.45\linewidth]{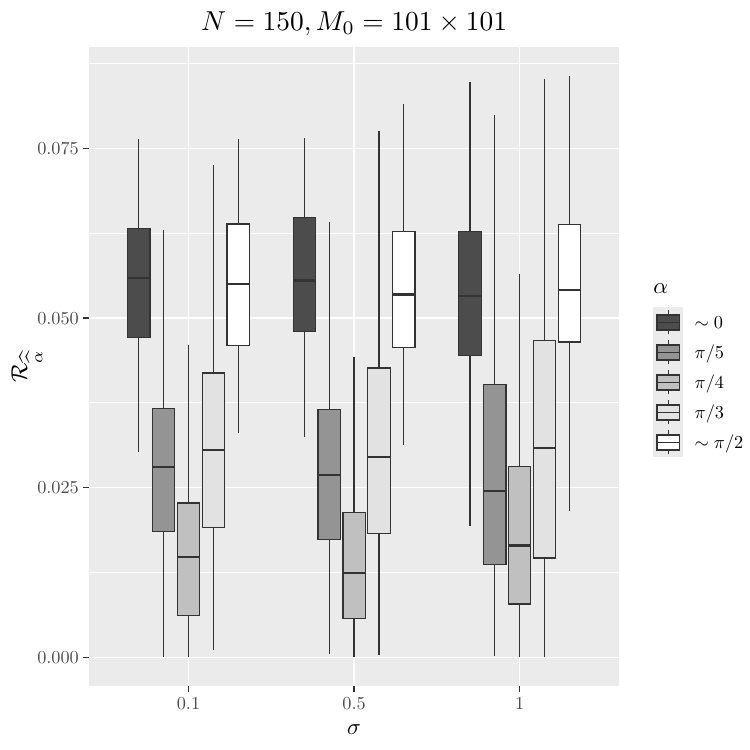} 
\end{minipage} 
\caption{Results for the risk of estimated angles (with correction). $N$ is the number of curves, $M_0$ the number of observed points along each curve, and $\sigma$ is the noise level.}
\label{fig:alpha-adj-box}
\end{figure}

Boxplots for the simulation results can be seen in Figure \ref{fig:alpha-adj-box}. The risk can be seen to be very small, below 0.1 for all different configurations, even in the extremely high noise setting $\sigma = 1$. In fact, we see that our Algorithms are fairly robust to noise, in the sense that the risk increase less than proportionately to the noise levels. As seen in Section \ref{sec:appli}, the levels of accuracy observed in Figure \ref{fig:alpha-adj-box} is sufficient to achieve lower risk levels for applications such as smoothing.

\subsection{Real data application}\label{sec:real-data}

To illustrate our methodology, we analyze gridded weather observations from the Oregon State PRISM Climate Project \citep{prism2025,daly-2013}. The data set, accessed using the \textbf{R} package \texttt{prism} \citep{hart2015prism}, contains annual weather measurements such as temperature, precipitation and vapor pressure deficit on a 4km grid resolution of the contiguous United States. We restrict our attention to yearly precipitation levels ranging over the period 1900 to 2023 ($N = 124)$.

\begin{figure}[ht]
    \centering
    \begin{minipage}[b]{0.47\linewidth}
        \centering
        \includegraphics[width=\linewidth]{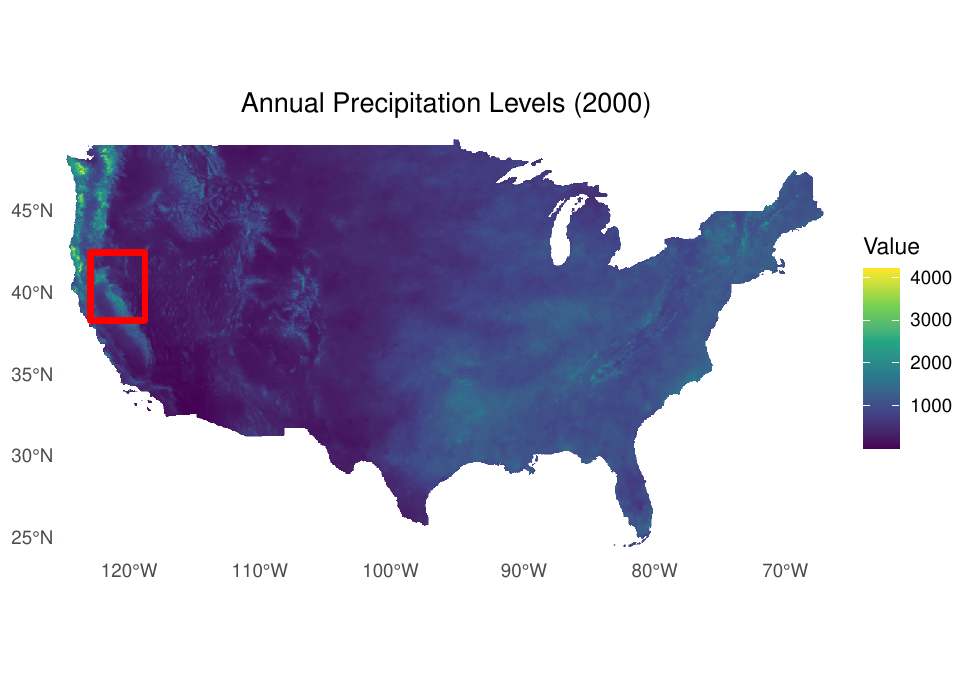}
        \caption{Rainfall (mm) of the Contiguous United States in 2000.}
        \label{fig:us-rainfall-2000}
    \end{minipage}
    \hfill
    \begin{minipage}[b]{0.49\linewidth}
        \centering
        \includegraphics[width=\linewidth]{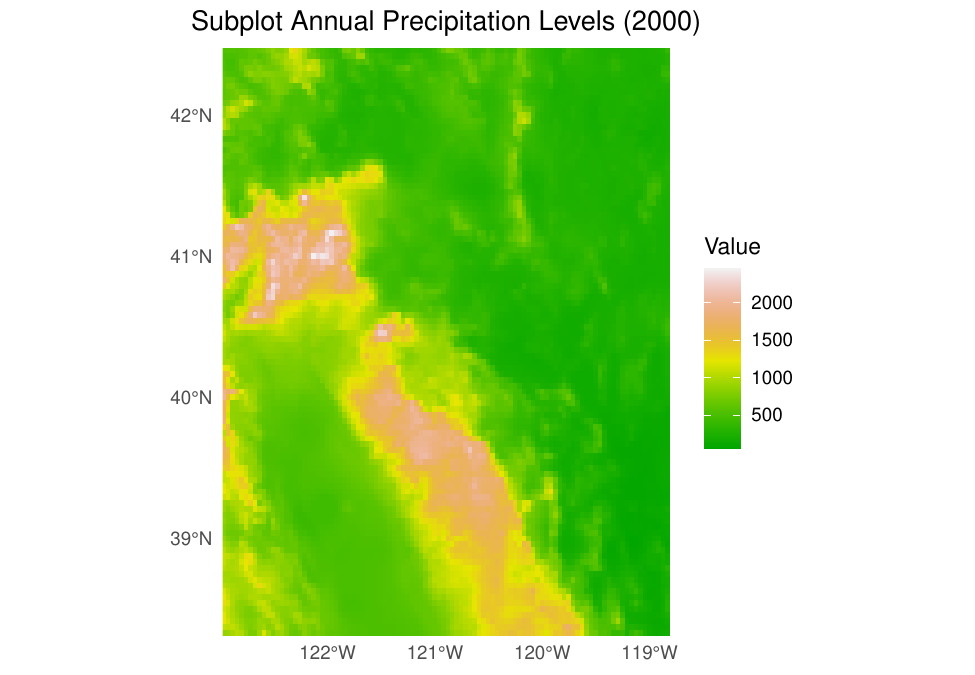}
        \caption{Rainfall (mm) of subplot (red box in Figure \ref{fig:us-rainfall-2000}) including the Northern Sierra Nevada Region in 2000.}
        \label{fig:ncal-rainfall-2000}
    \end{minipage}
\end{figure}

We treat the rainfall measurements $Y^{(j)}_m$ as two-dimensional functional data indexed by geographic coordinates $\mathbf{t} = (\text{longitude, latitude}) \in \mathcal{T}$, with temporal replicates indexed by $j = 1, \dots, 124$. An example of one replication can be seen in Figure \ref{fig:us-rainfall-2000}. Given the United States’ varied topography, it is unsurprising that anisotropy varies with location. Although our methodology is tailored to the case of constant anisotropy over the domain $\mathbf{t} \in \mathcal{T}$, one can readily adapt it via localization. As an illustrative example, we isolate and work with a $100 \times 100$ subgrid (i.e., $M_0 = 100^2$) which encompasses the Northern Sierra Nevada Region, spanning approximately $122.98^\circ$ to $118.81^\circ$ W longitude, and from $38.31^\circ$ to $42.48^\circ$ N latitude. This region is likely to exhibit anisotropic precipitation patterns due to orographic effects across the Sierra Nevada and California Coast Ranges. Indeed, visual inspection of Figure \ref{fig:ncal-rainfall-2000} suggests anisotropy with an angle close to $3\pi/4$.

Each coordinate in $\mathcal{T}$
was normalized to lie within the unit square by subtracting its minimum and dividing by its range. Algorithm \ref{algo:alpha-hat-algo} was then applied to estimate the angle $\widehat \alpha$. For the estimation of $ \widehat{g_{\alpha}}( \Delta)  $, we select the spacing parameter $\Delta = \exp(-\log^{0.4}(M_0))$, while for the identification process, we use an equally spaced grid of 21 points, with $\boldsymbol \Delta = \{\exp(-\log^{0.4}(M_0)), \dots, 0.4\}$.

The angle estimates are reported in Table \ref{tab:transposed-result}. We see that the identified angle corresponds to $g = -\cot$, with an estimate $\widehat \alpha = 2.24$. This aligns closely with the visual result of an angle close to $3\pi/4$.

\begin{table}[htbp]
\centering
\caption{\centering Algorithm Results}
\label{tab:transposed-result}
\begin{tabular}{cccc}
\toprule
$g$ & $\widehat \alpha$ & $\widehat{\underline{H}}$ & $\widehat{\overline{H}}$ \\
\midrule
$-\cot$ & $2.24$ & $0.34$ & $0.37$ \\
\bottomrule
\end{tabular}

\end{table}

\section{Applications}\label{sec:appli}
In this section, we discuss concrete applications of our methodology. They serve as a strong motivation for performing a change of basis as a standard pre-processing step. 

\subsection{Smoothing bivariate functional data}
Let $\{X(\mathbf{t}), \mathbf{t} \in \mathcal{T} \subset \mathbb{R}^2 \}$ be a bi-variate stochastic process satisfying \eqref{eq:mean-squared-variation}, with maximizing regularity along the  direction $\mathbf{u}_1 = \cos(\alpha)\mathbf{e}_1 + \sin(\alpha)\mathbf{e}_2$. Following the preceding sections, our exposition will be centred around the common design case. Observations come in the form of pairs $\mathcal{D}_0 = (Y_m^{(j)},\mathbf{t}_m), 1 \leq m \leq M_0, 1 \leq j \leq N$, generated under
\begin{equation*}
	Y^{(j)}_m = X^{(j)}(\mathbf{t}_m) +   \sigma e^{(j)}_m, \qquad 1\leq m\leq M_0, 1 \leq j \leq N,
\end{equation*}
where $e_m^{(j)}$ are i.i.d centered random variables with unit variance. We call $\mathcal{D}_0$ the learning set. 

Consider a new realization $X^{new}$ of $X$, where the observed pairs $\mathcal{D}_1 = (Y^{new}_m,\mathbf{t}_m), 1\leq m\leq M_0$ are generated from 
\begin{equation*}
	Y^{new}_m=X^{new}(\mathbf{t}_m)+\sigma e^{new}_m,\quad \quad 1\leq m\leq M_0,
\end{equation*}
where $e_m^{new}$ are i.i.d centered random variables with unit variance, and $X^{new}$ and $e_m^{new}$ are independent. We refer to $\mathcal{D}_1$ as the online set. 

Our goal is the recovery of $X^{new}$ with $\mathcal{D}_0$, using a suitable estimator $\widehat{X}^{new}$. Several methodologies for smoothing multivariate functional data currently exist. For example, \citep{ramsay2002, sangalli2013, wood2008} consider spatial regression using a penalty that involves the Laplacian. These do not take the anisotropy of the process into account. \citep{azzim2015} and \citep{bernardi2018} extends these previous approaches by using a penalty term involving the partial differential operator, taking the anisotropy of the process into account. All these methods assume that the processes are differentiable for the penalty to be well defined. This assumption does not always hold in practice, and we propose an approach which works for non-differentiable processes.


Let $R_\alpha$ be a clockwise rotation matrix given by
\begin{equation*}
R_{\alpha} = \begin{pmatrix}
\cos(\alpha) & \sin(\alpha) \\
-\sin(\alpha) & \cos(\alpha)
\end{pmatrix}.
\end{equation*}
Define a new process by applying the rotation matrix to the sampling points, denoted by
\begin{equation*}
Z(\mathbf{t}) := X(R_{\alpha}^{-1} \cdot \mathbf{t}), \qquad \forall \mathbf{t} \in \mathcal{T}.
\end{equation*}
The process $Z(\mathbf{t})$ satisfies \eqref{eq:mean-squared-variation} with maximizing regularity along the canonical direction $\mathbf{e}_2$.
It is thus beneficial to work with $Z$ instead of $X$ on the canonical basis. In practice, this transformed process can be obtained by performing a change-of-basis after estimating the rotation matrix $R_{\alpha}$.

For simplicity, we fix the smoother to be the Nadaraya-Watson estimator with the multiplicative Epanechnikov kernel $K:\mathbb R ^2\to \Rplus$, supported on $[-1,1]\times [-1,1]$, where
\begin{equation*}
K(\mathbf s)= K_{ep}(s_1) \times K_{ep}(s_2), \quad \forall \mathbf s=(s_1,s_2)\in \mathbb R^2, \quad  \text{and } K_{ep}(x)= \frac{3}{4}(1-x^2)\mathbb{I}_{\left\{|x|\leq 1\right\}}.
\end{equation*}
Furthermore, $\mathbf{B}=\operatorname{diag}(h_1^{-1}, h_2^{-1})$ is a positive definite bandwidth matrix. Using the rule $0/0=0$, the Nadaraya-Watson estimator is given by
\begin{equation*}
\widehat Z^{new}(\mathbf{t};\mathbf{B} )=\sum_{m=1}^{M_0}Y^{new}_m\frac{K\left(\mathbf{B}(R_\alpha\mathbf{t}^{new}_m-\mathbf{t})\right)}{\sum^{M_0}_{  m ^{\prime}   =1} K\left(\mathbf{B}(R_\alpha\mathbf{t}^{new}_{ m^{\prime}  } -\mathbf{t})\right)}. 
\end{equation*}
The bandwidths should adapt to the regularity of the process $Z$ to obtain the optimal rate of convergence. In particular, the bandwidths should be selected adaptively to the intrinsic anisotropy of the process. This can be achieved through a plug-in bandwidth rule when explicit risk bounds are available and directly estimable from the data. The estimate of $X^{new}$ is then obtained  by
\begin{equation*}
\widehat X^{new}(\mathbf t, \mathbf B)= \widehat Z^{new}(R_{\alpha}\mathbf t, \mathbf B), \qquad \forall \mathbf t\in \mathcal T.
\end{equation*}
Let $B(\mathbf 0,r)$ denote the ball centered at the origin of $\mathbb R^2$ with radius $r$. Consider the $L^2$ risk of $\widehat Z^{new}$:
\begin{equation}\label{eq:ave_risk}
\mathcal R \left(\textbf B, M_0\right)=\mathbb{E}\left[\|\widehat Z ^{new}(\cdot\hspace{0.1cm};\textbf B)-Z^{new}(\cdot)\|_{2}\right]=\mathbb{E}\left[\|\widehat X ^{new}(\cdot\hspace{0.1cm};\textbf B)-X^{new}(\cdot)\|_{2}\right],
\end{equation}
since the matrix $R_\alpha$ has determinant 1. 
Proposition \ref{prop:risk-smooth} provides a $L^2$ risk bound.

\begin{proposition}\label{prop:risk-smooth}
Let $h_1>0$, $h_2>0$ be in a bandwidth range satisfying
\begin{equation}\label{eq:bth-range}
\max\{h_1, h_2\} \rightarrow 0, \qquad \text{and} \quad \sqrt{M_0}\times \min \{h_1,h_2\} \rightarrow \infty.
\end{equation}
Then the following risk bound holds: 
\begin{equation}\label{eq:risk_1}
	\mathcal{R}(\textbf B,M_0) \lesssim \left\{ \frac{1}{M_0 h_1 h_2}+ h_1^{2H_1} + h_2^{2H_2} \right\}.
\end{equation}
\end{proposition}
The proof is provided in the Supplementary Material \citep{kassiwangsupp25}. The following corollary provides the optimal bandwidth with respect to the $L^1$ norm. 
\begin{corollary}\label{cor:opt-h-smooth}
Under the assumptions of Proposition \ref{prop:risk-smooth}, the optimal bandwidths $h_1^*$ and $h_2^*$ satisfies
\begin{equation}\label{bandchoice}
h_1^* \asymp \left(\frac{1}{M_0}\right)^{\frac{H_2}{2H_1H_2 +H_1+ H_2}}, \quad  h_2^* \asymp \left(\frac{1}{M_0}\right)^{\frac{H_1}{2H_1H_2 +H_1+ H_2}}.
\end{equation}
Moreover, if $h_1^*$ and $h_2^*$ is used for smoothing, the following rate is obtained: 
\begin{equation*}
\mathcal{R}( \textbf B^*,M_0) \lesssim  M_0^{-\frac{2\omega}{2\omega+1}},
\end{equation*}
where $\omega$ is the "effective smoothness", defined by the relation 
\begin{equation*}
\frac{1}{H_1}+\frac{1}{H_2} = \frac{1}{\omega}.
\end{equation*}
\end{corollary}
All the quantities in Corollary \ref{cor:opt-h-smooth} are estimable from the data using our methodology. Structural adaptation enables the optimal bandwidths to be chosen according to Corollary \ref{cor:opt-h-smooth}, obtaining the anisotropic rates of convergence when the intrinsic anisotropy is not in the direction of the canonical basis. 
 
Let $\widehat{h}^*_1$ and $\widehat{h}^*_2$ denote the bandwidths obtained as in \eqref{bandchoice}, where $H_1$ and $H_2$ are replaced by their respective estimates. It is worth noting that the rotation matrix $R_{\widehat{\alpha}}$ has no effect on the integrated risk $\mathcal{R}(\widehat{\mathbf{B}}^*, M_0)$, as its determinant is equal to 1. The only influence of the estimated angle $\widehat{\alpha}$ lies in the estimation of $\widehat{H}_2 = \widehat{H}_{\mathbf{u}(\widehat{\alpha})}$.
Furthermore, we only require a convergence rate of order $\log^{-1}(M_0)$ for the estimation of the regularities. This is justified by the identity $M_0^{1/\log(M_0)} = \exp(1)$, which implies that such a rate is sufficient. To achieve this rate, the learning set $\mathcal{D}_0$ must be sufficiently large. 
\begin{proposition}
    Suppose the conditions of Proposition \ref{g:concentration} and Theorem \ref{alpha:concen} hold. Moreover, assume that $K_0 = \log(M_0)$ and $\Delta = \exp(-\log^\xi(M_0))$ for some $0 < \xi < 1$, and that the following condition is satisfied: there exists $b>0$ such that 
\begin{equation}\label{Nlarge_mean}
    \frac{M_0^{b}}{N_0} \lesssim \log^{2+2\xi}(M_0) \exp\left( 4(H_1 + H_2)\log^{\xi}(M_0) \right).
\end{equation}
Then, it holds that
\begin{equation*}
  \mathcal{R}( \widehat{\mathbf B}^*,M_0)   \lesssim  M_0^{-\frac{2\omega}{2\omega+1}}.
\end{equation*}
\end{proposition}

By working only on the canonical basis, the rates of convergence that is generally obtained corresponds to the isotropic rate. This is confirmed by our simulation study, which we discuss now. 

Our simulation study aims to compare the $L^1$ risk of smoothing, with and without a change of basis. Surfaces corresponding to the sum of fBms with regularities $H_1 = 0.8$, $H_2 = 0.5$ were simulated using the Algorithm described in Section \ref{sec:simulator}. The angles $\alpha$ were given by $\alpha \in \{\pi/30, \pi/6, \pi/4, \pi/3, \pi/2 - \pi/30\}$. $N = 120$ surfaces were generated with $M_0 = 101^2$ evenly spaced sampling points. Gaussian noise $\sigma e_m^{(j)}$ were added, where $e_m^{(j)} \stackrel{iid}{\sim}\mathcal{N}(0, 1)$ and $\sigma \in \{0.1, 0.5, 1 \}$. Estimates of $\alpha$ were obtained with Algorithm \ref{algo:alpha-hat-algo}, with $\Delta = M_0^{-1/4}$, $\boldsymbol{\Delta} = \{M^{-1/4}, \Delta_1, \dots, \Delta_{K_0 - 1}, 0.4 \}$, and $\#\boldsymbol{\Delta} = 15$.

The online set, consisting of one surface per replication, was generated from the same process, for a total of 400 replications. The true surface was first generated on an equally spaced grid of $201^2$ points on $\mathcal{T}$ without noise. Nearest-neighbor interpolation was then performed to discretize the process onto a grid of $101^2$ points. The same Gaussian noise as the learning set was added.

For the anisotropic setup, a change of basis is performed on the online set. The minimum and maximum regularities were estimated using \eqref{eq:H-min-estim} and \eqref{eq:H_estim_dir}. The bandwidths were selected according to Corollary \ref{cor:opt-h-smooth}. In the isotropic setup, no change of basis is performed, and the bandwidths were similarly selected according to Corollary \ref{cor:opt-h-smooth}, with $H_1 = H_2$. A multiplicative kernel was used, where the Epanechnikov kernel was used in each dimension. 

\begin{figure}[!ht]
\centering
\includegraphics[height = 0.3\textheight,width=.5\textwidth]{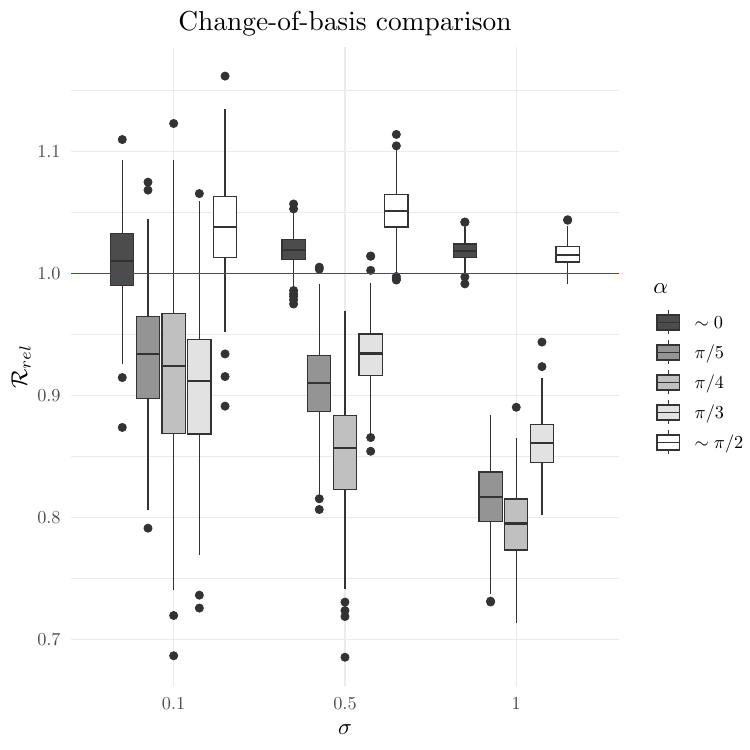}
\caption{Results for the relative risk of smoothing surfaces with and without a change-of-basis for the different angles $\alpha$ at noise level $\sigma$.}
\label{fig:smooth-box}
\end{figure}

The risk measure was taken to be the relative risk, given by
\begin{equation*}
\mathcal{R}_{rel} = \frac{\mathcal{R}_{ani}}{\mathcal{R}_{iso}},
\end{equation*}
where $\mathcal{R}_{ani}$ and $\mathcal{R}_{iso}$ correspond to the $L^2$ anisotropic and isotropic risk respectively. Simulation results can be seen in Figure \ref{fig:smooth-box}. We see that with the exception of the angles near the boundary (i.e., 0 or $\pi/2$), performing a change of basis leads to a reduction in the $L^2$ risk, by levels as much as 10\%. It is not surprising that at the boundary, the risk is worse, since one is already anisotropic along the canonical basis. 

\subsection{Anisotropic detection}\label{sec:aniso-detect}
Our focus so far has been on estimating the direction of the maximizing regularity, which implicitly assumes that anisotropy is present. When the process is intrinsically isotropic, no gains can be made by structural adaptation since the regularity of the process $X$ is invariant to directions. However, there is no real loss in terms of statistical error either, since the underlying regularity remains the same, as implied by Lemma \ref{lm}. In particular, the rates of convergence do not get degraded, and can only be improved with structural adaptation. 

Nevertheless, there might be instances where knowing the presence of anisotropy is relevant for some applications. Some examples include fingerprint verification \citep{jainhong1997,jiang2005}, and texture analysis in materials science \citep{GERMAIN2003}. This issue can be naturally addressed using the directional regularity approach, by constructing an anisotropic detection procedure based on thresholding. The underlying idea is to consider an event 
\begin{equation*}
\mathcal{A}(\tau) := \left\{\left|\widehat{\underline{H}} - \widehat{\overline{H}}\right| > \tau \right\},
\end{equation*}
and determine anisotropy based on $\mathbb{I}_{\mathcal{A}(\tau)} $, for some threshold $\tau$ that is appropriately chosen. In particular, $\tau$ should be selected to account for the estimation errors of $\widehat{\underline{H}}$ and $\widehat{\overline{H}}$.

Let $\underline \varepsilon = |\widehat{\underline{H}} - \underline{H}|$ and $\overline \varepsilon = \overline{H}- \underline{H} $ denote the estimation error of the minimum regularity and the difference of the true regularities respectively. Under anisotropy, for $M_0$ and $N$ sufficiently large, we have $\overline \varepsilon \geq \underline \varepsilon$ with high probability. On the one hand, when $\underline H = \overline H$ (i.e., isotropy), choosing $\tau > \underline \varepsilon$  accounts for the estimation error so that $\mathbb{I}_{\mathcal{A}(\tau)} = 0$ with high probability. On the other hand, when $\underline H \neq \overline H$, choosing $\tau < \overline \varepsilon$ is necessary to avoid falsely detecting isotropy. An illustration of this can be seen in Figure \ref{fig:tau-illu-line} for the case of anisotropy. 

\begin{figure}
\centering
\begin{tikzpicture}
\draw(0,0)--(10,0);
\foreach \x/\xtext in {0/0, 2/$\underline \varepsilon$, 2.5/$\tau$, 4/$\overline \varepsilon$, 10/1}
    \draw(\x,5pt)--(\x,-5pt) node[below] {\xtext};
\draw[decorate, decoration={brace}, yshift=2ex]  (0,0) -- node[above=0.4ex] {Always detect anisotropy}  (2,0);
\draw[decorate, decoration={brace, mirror}, yshift=2ex]  (10,0) -- node[above=0.4ex] {Always detect isotropy}  (4,0);
\end{tikzpicture}
\caption{Illustration of the region where the thresholding parameter $\tau$ should fall under anisotropy ($\overline H \neq \underline H $).}
\label{fig:tau-illu-line}
\end{figure}

Since $\underline \varepsilon$ is unknown in practice, we construct a data-driven procedure to estimate it, which we denote $\widehat{\underline{\varepsilon}}$. Let $\beta = \{\beta_1, \dots, \beta_J \}, 0 < J < \infty$ be a random set of angles, where $J = J(N, M_0)$ is an integer that depends on the sample size. Let $\beta^{\perp} =  \{\beta_1^{\perp}, \dots, \beta_J^{\perp}\}$ be the set of angles orthogonal to $\beta$, where $\beta_j^{\perp} = \beta_j + \pi/2$, $\forall j = 1, \dots, J$. The main idea is to estimate the minimum regularity associated with each $\beta_j$ and $\beta_j^{\perp}$, and compute the average difference between these two estimates, given by

\begin{equation}\label{eq:H-min-err}
\widehat{\underline{\varepsilon}} = \frac{1}{J}\sum_{j=1}^{J} \underline \varepsilon_j,
\end{equation}
where for all $j = 1, \dots, J$, we have
\begin{equation}\label{eq:H-min-err-k}
\underline \varepsilon_j = \left|\check H_{\mathbf{u}(\beta_j)} - \check H_{\mathbf{u}(\beta_j^{\perp})} \right|, \text{ and } \check{H}_{\mathbf{u}(\beta)} = \frac{1}{\#\boldsymbol{\Delta}}\sum_{k=1}^{K_0}  \widehat{H}_{\mathbf{u}(\beta)}(\Delta_k).
\end{equation}
The principle for this approach is rooted in Lemma \ref{lm}, which states that under anisotropy, there exists only one maximizing direction $\mathbf{u}(\alpha)$. Thus if we take any other random direction $\mathbf{v}(\beta)$, $\beta \neq \alpha$, then the regularity that we will "catch" corresponds to the minimum one.

The quantity $\widehat{\underline{\varepsilon}}$ in  \eqref{eq:H-min-err} estimates the "irreducible error" that arises in estimating the worst regularity $\underline H$ by simply changing directions, exemplified by Proposition \ref{prop:continuity_proxy}. In order to avoid always detecting anisotropy, this error needs to be taken into account, so that $\tau > \underline \varepsilon$. Averaging the estimated regularities over a grid of $\boldsymbol{\Delta}$'s provides added robustness since $\widehat{H}_{\mathbf{u}(\beta)}(\Delta_k)$ is no longer simply an auxiliary quantity.

There are two reasons for computing the $\underline \varepsilon_j$'s using different groups of estimates in $\beta$ and $\beta^{\perp}$. The first is to preserve the independence between summands when computing $\widehat{\underline{\varepsilon}}$ in \eqref{eq:H-min-err}. The second is to ensure that the angles are sufficiently well separated to ensure that the difference in the estimates are not driven by the proximity of angles. The threshold is then set to 
\begin{equation}\label{eq:tau-hat}
\tau = \widehat{\underline{\varepsilon}} + \exp(-\log^\xi(M_0)),
\end{equation}
for some $0 < \xi < 1$. The extra $\exp(-\log^\xi(M_0))$ term in \eqref{eq:tau-hat} converges to zero slower than the rate of convergence of $\widehat{\underline{H}}$, and ensures that the strict inequality $\underline \varepsilon < \tau < \overline \varepsilon$ is preserved for any  $M_0$ large enough. Algorithm \ref{algo:ani-detect} provides a summary of the anisotropic detection procedure.

\begin{algorithm}
\caption{Anisotropic Detection}
\label{algo:ani-detect}
\begin{algorithmic}[1]
\Require Data $(Y^{(j)}(\mathbf{t}_m), \mathbf{t}_m)$, Grid $\widetilde{\mathcal{T}} = \left\{\mathbf{t_1}, \dots, \mathbf{t_p} \right\}$; Integer $J$; Estimated Angle $ \check{\alpha}$

Initialize $H \gets \emptyset$;

\For{{$j = 1, \dots, J$}}

\State Sample $\beta_j \sim \text{Unif}([ \check{\alpha}    + \pi/4,   \check{\alpha}    + 3\pi/4])$;

\State Estimate $\widehat H_j = \widehat{H}_{\mathbf{v}(\beta_j)}$ according to \eqref{eq:H_estim_dir};

\State $H \gets \widehat H_j \bigcup H$;

\EndFor

\State Compute $\widehat{\underline{\varepsilon}}$ according to \eqref{eq:H-min-err};

\State Compute $\tau$ according to \eqref{eq:tau-hat};

\State \Return $\mathbb{I}_{\mathcal{A}(\tau)}$;
\Comment 1 indicates anisotropy, isotropy otherwise

\end{algorithmic}
\end{algorithm}
In theory, the set of angles $\mathbf{\beta}$ can be chosen randomly, as long as $\mathbf{\beta} \neq \pm   \check{{\alpha}}$. However, in order to avoid any "continuity" issues, each $\beta_j$ should be sufficiently far away from $ \check{\alpha}$. We thus suggest to sample the angles from a uniform distribution, as seen in Algorithm \ref{algo:ani-detect}. The power $\xi$ in \eqref{eq:tau-hat}, which governs the rate of convergence, only needs to be such that $\xi \in (0,1)$ in theory. Following \citep{Golovkine2022}, we choose $\xi = 1/3$, a value that seems to work well in practice. The integer $J(N, M_0)$ should be increasing with $N$ and $M_0$, so that $\widehat{\underline{\varepsilon}}$ eventually converges. In view of computational efficiency, we suggest to select $J(N, M_0) = \lceil (N \times M_0)^{1/4} \rceil$, which is aligned with the rate of convergence of $\widehat H$ and also works well in practice. 

We establish the consistency of Algorithm \ref{algo:ani-detect} in the following proposition.
\begin{proposition}\label{prop:ani-detect-consis}
Let $\widehat{\overline H} = \widehat H_{\mathbf{u}(\widehat \alpha)}$, $\widehat{\underline H} = \widehat H_{\mathbf{u}(\widehat \alpha + \pi/2)}$, and $\tau$ be defined as in Algorithm \ref{algo:ani-detect}. Suppose that the assumptions of Corollary \ref{para:setting} hold true. Then Algorithm \ref{algo:ani-detect} is consistent, in the sense that 
\begin{equation*}
\lim_{N, M_0 \rightarrow \infty} \mathbb{P}\left(\left|\widehat{\overline{H}} - \widehat{\underline{H}} \right| > \tau, \underline H \neq \overline H \right) = 1, \quad  \text{and} \quad \lim_{N, M_0 \rightarrow \infty} \mathbb{P}\left(\left|\widehat{\overline{H}} - \widehat{\underline{H}} \right| > \tau, \underline H = \overline H \right) = 0.
\end{equation*}
\end{proposition}
The proof can be found in the Supplementary Material \citep{kassiwangsupp25}. Simulation results for the anisotropic detection procedure described in Section \ref{sec:aniso-detect} can be seen in Table \ref{table:ani-detect}. The surfaces were generated in a similar manner as before, using the sum of fBms. The following parameter choices were made: $\alpha = \pi/3$, $\underline H = 0.5$, $\sigma = 0.1$, with 500 replications. The percentage column indicates the percentage of cases classified as anisotropic ($\mathbb{I}_{\mathcal{A}} = 1$). In the isotropic case, we achieve virtually perfect classification, even for relatively small values of $M_0$. In the anisotropic case, we require either a sufficiently large number of points observed on each surface, or for the difference $\overline H - \underline H$ to be well-separated. When the difference in regularities is large enough, we can achieve almost perfect classification for $M_0 = 101 \times 101$. In the context of regularity estimation and anisotropic detection, this is a relatively small number of points. For example, \citep{richard2016test} constructs a statistical test for isotropic detection, where the number of surfaces $N$ plays a more important role. In his simulations, $N = 6000$, a much larger quantity.

\begin{table}[]
\centering
\begin{tabular}{@{}lllll@{}}
\toprule
Setup       & $\overline H$ & $\sigma$ & $M_0$  & Percentage \\ \midrule
Isotropic   & 0.5      & 0.1   & $51 \times 51$  & 0     \\
Isotropic   & 0.5      & 0.1   & $101 \times 101$ & 0     \\
Anisotropic & 0.8      & 0.1   & $51 \times 51$  & 35.8  \\
Anisotropic & 0.8      & 0.1   & $101 \times 101$ & 42.4  \\
Anisotropic & 0.9      & 0.1   & $51 \times 51$  &  80.6 \\
Anisotropic & 0.9      & 0.1   & $101 \times 101$ & 97    \\ \bottomrule
\end{tabular}
\caption{Table showing the results of the anisotropic detection approach using thresholding. The percentage column indicates the percentage of cases classified as anisotropic (i.e., $\mathbb{I}_{\mathcal{A}} = 1$) for the different setups.}
\label{table:ani-detect}
\end{table}

\section{Conclusion and discussion}\label{sec:discussion}
In this section we   conclude by discussing some key aspects of our methodology and future directions. 

\subsection{Computational considerations}\label{sec:computational}
The computational complexity of Algorithm \ref{algo:alpha-hat-algo} is driven primarily by the identification process, due to the additional loop over the grid of spacings $\boldsymbol{\Delta}$. Since we are performing a linear grid search in each basis vector containing $\sqrt{M_0}$ points, the complexity of nearest-neighbor interpolation for each surface is $O(M_0)$, resulting in $O(N \times M_0)$ for $N$ sample paths. By searching over $M_0$ points to determine the closest point to each $\mathbf{t}_m$ in terms of $\ell_1$ norm, the complexity of computing $\widehat \sigma^2$ is $O(N \times M_0)$, so the complexity of $\widehat H_{\mathbf{v}(\beta)}(\mathbf{t}, \Delta)$ is $O(N \times M_0)$ for each $\mathbf{t}$, $\Delta$. The complexity of the full identification process is therefore $O(N \times \#\widetilde{\mathcal{T}} \times K_0 \times M_0)$. In the context of fda, a complexity of at least $O(N \times \#\widetilde{\mathcal{T}} \times M_0)$ is to be expected. Since $K_0$ is relatively small (e.g., 15 points), our algorithm is reasonably good in terms of computational complexity, considering the possible gains in the rates of convergence.

Clearly, the simulation procedure described in Section \ref{sec:simulator} depends on the method used to generate the individual fBms. Once the fBms are given, only a linear search over $\#\widetilde{\mathcal{T}}$ grid points is required. Using a fast simulation method  such as the circulant embedding method in \citep{woodchan94}, individual one dimensional curves can be simulated in $O(\{\#\widetilde{\mathcal{T}} \times \log(\#\widetilde{\mathcal{T}})\}^{1/2})$ time. Thus the simulation of $N$ surfaces can be done in $O(N \times  \#\widetilde{\mathcal{T}} \times \log(\#\widetilde{\mathcal{T}}))$. To the best of our knowledge, we do not know of any simulation algorithm in the context of fda that has better computational complexity. 

\subsection{Singularity along the maximizing direction}\label{sec:singular-dis}
Lemma \ref{lm} states that there is a singularity present in the concept of directional regularity, in the sense that there exists only one direction for which the regularity is maximal. Since the angles $\alpha$ are not expected to be estimated perfectly, why can there be a gain in convergence rates, as seen in the empirical results in Sections \ref{sec:emp-results}?

The answer lies in the intrinsic nature of directional regularity, which is itself intimately related to the mean-squared variations of a process. It is an "infinitesimal" concept, in the sense that
\begin{equation}
 \lim_{\Delta \rightarrow 0}\frac{\theta_{\mathbf{u}}(\mathbf{t}, 
 \Delta ))}{L_{\mathbf{u}}(\mathbf{t}) \Delta^{2H_{\mathbf{u}}}} = 1.
\end{equation}
However, the limiting case cannot be obtained with finite precision computers. In reality, the "true directional regularity" that one observes is given by the proxy $H_{\mathbf{u}(\beta)}(\Delta)$. This partly explains the observed "continuity" associated with the directional regularity in practice, so that the   estimation error    of $\alpha$ only needs to be sufficiently small in order for gains in the rate   of convergence    to be achieved. 

A more precise, formal perspective can be taken through the lens of Proposition \ref{prop:continuity_proxy}. In the upper bound of $H_{\mathbf{u}(\beta_1)} (\Delta)   - H_{\mathbf{u}(\beta_2)} (\Delta)  $, we observe two competing terms, in the form of $\Delta^{2(\underline H - H_{\mathbf{u}(\beta_{ k   })})}$ and $(\beta_1 - \beta_2)^{2\underline H}$. In order for $H_{\mathbf{u}(\beta_1)}(\Delta) - H_{\mathbf{u}(\beta_2)}(\Delta)$ to converge as $\beta_1 \rightarrow \beta_2 $, it is necessary that the condition
\begin{equation}\label{eq:conti-cond}
     (\beta_1 - \beta_2)^{2\underline H}= o\left( \Delta^{2(H_{\mathbf{u}(\beta_{k  })} - \underline H)} \right)  
\end{equation}
is satisfied. Equation \eqref{eq:conti-cond} can be loosely interpreted as saying "for directions that are close enough to each other, the proxies in \eqref{eq:H_estim_dir}, of the  directional regularity stays somewhat close"; see Proposition \ref{prop:continuity_proxy}.

\subsection{Higher dimensional domains} 
Although the main focus of this paper is on the bivariate case $\mathcal T \subset \mathbb R^2$, Definition~\ref{def:direg-def} naturally extends to higher-dimensional domains. The main conceptual and methodological challenges are highlighted. 

Let $\mathcal T \subset \mathbb R^d$ with $d\geq 3$ and define the unit sphere as $\mathbb S^{d-1}= \{\mathbf u \in \mathbb R^d: \|\mathbf u\|=1\}$. For a second-order stochastic process $\{X(\mathbf t), \mathbf t \in \mathcal T\}$, directional regularity can be defined analogously by considering mean-squared increments along directions $\mathbf u \in \mathbb S^{d-1}$:
$$
\theta_{\mathbf u}(\mathbf{t}, \Delta):=\mathbb{E}\left[\left\{  X\left( \mathbf{t} - \frac{\Delta}{2} \mathbf{u} \right)- X\left( \mathbf{t} + \frac{\Delta}{2}\mathbf{u} \right) \right\}^2 \right] = L_{\mathbf{u}}(\mathbf{t}) \Delta^{2H_{\mathbf{u}}(\mathbf{t})} + G_{  \mathbf{u}   } (\mathbf{t},\Delta), 
$$
where $H_{\mathbf u}(\mathbf t) \in (0,1)$ denotes the local regularity along the direction $\mathbf u$. For the sake of clarity, we assume that for any direction $\mathbf u \in \mathbb S^{d-1}$, the local regularity $H_{\mathbf u} $ is constant across the domain $\mathcal T$.  

The geometry of $\mathbb S^{d-1}$ allows for a richer structure of anisotropy. In contrast to the bivariate case, maximal directional regularity need not be achieved along a single direction, but may instead correspond to an entire subspace.

More precisely, one may postulate the existence of an orthogonal decomposition
$$
\mathbb{R}^d = E_1 \oplus \cdots \oplus E_K,
\qquad
H_1 < \cdots < H_K, \quad  K\leq d,
$$
where each subspace $E_k$ is associated with a regularity exponent $H_k \in (0,1)$. The restriction of the process $X$ to the subspace $E_k$ is an isotropic process with regularity $H_k$. Under a suitable expansion of the directional mean-squared increments, the regularity along a direction $\mathbf v \in \mathbb S^{d-1}$ is then determined by the least regular subspace onto which $\mathbf v$ has a non-zero projection. In other words,
$$
H_{\mathbf v}
=
\min\; \left\{ H_k : \Pi_{E_k} (\mathbf v) \neq 0 \right \},
$$
where $\Pi_{E_k}$ denotes the orthogonal projection onto $E_k$. In particular, the maximal regularity $H_K$ is attained if and only if $\mathbf u$ belongs to the most regular subspace $E_K$. The bivariate dichotomy of Lemma~\ref{lm} corresponds to the special case of $d=2=K$ and  $\dim(E_K) = 1$, in which only two antipodal directions ($\mathbf u$ and $-\mathbf u$) exhibit maximal regularity.

This suggests that in higher dimensions, the identification of anisotropy amounts to estimating a lower dimensional regularity subspace rather than a single direction. While a complete treatment of this setting is beyond the scope of the present work, the bivariate framework developed here can be viewed as the fundamental building block for such higher-dimensional extensions.

\begin{acks}[Acknowledgments]
We thank Valentin Patilea for his careful reading and providing detailed suggestions. We thank the Editors and two anonymous referees for constructive suggestions which greatly improved the paper.
\end{acks}

\begin{funding}
The authors acknowledge the support of the French Agence Nationale de la Recherche (ANR) under reference ANR-24-CE40-2439 (FUNMathStat  project). Sunny Wang also acknowledges funding within
the framework of the France 2030 programme for EUR DIGISPORT (ANR-18-EURE-0022) project.
\end{funding}

\section*{Supplementary Material}
In the supplement, detailed proofs and additional simulation results are provided.

\bibliographystyle{imsart-nameyear} 
\bibliography{biblio_final_cleaned}       

@article{BELLONI2015,
AUTHOR = {Belloni, Alexandre and Chernozhukov, Victor and Chetverikov,
Denis and Kato, Kengo},
TITLE = {Some new asymptotic theory for least squares series: pointwise
and uniform results},
JOURNAL = {J. Econometrics},
FJOURNAL = {Journal of Econometrics},
VOLUME = {186},
YEAR = {2015},
NUMBER = {2},
PAGES = {345--366}
}

@article{Golovkine2021,
    AUTHOR = {Golovkine, Steven and Klutchnikoff, Nicolas and Patilea,
              Valentin},
     TITLE = {Adaptive estimation of irregular mean and covariance
              functions},
   JOURNAL = {Bernoulli},
  FJOURNAL = {Bernoulli. Official Journal of the Bernoulli Society for
              Mathematical Statistics and Probability},
    VOLUME = {31},
      YEAR = {2025},
    NUMBER = {2},
     PAGES = {1032--1057},
      ISSN = {1350-7265,1573-9759},
   MRCLASS = {62G05 (62R10)},
  MRNUMBER = {4863066},
       DOI = {10.3150/24-bej1759},
       URL = {https://doi.org/10.3150/24-bej1759},
}

@book {Ramsay2005,
    AUTHOR = {Ramsay, J. O. and Silverman, B. W.},
     TITLE = {Functional Data Analysis},
    SERIES = {Springer Series in Statistics},
   EDITION = {Second},
 PUBLISHER = {Springer, New York},
      YEAR = {2005},
     PAGES = {xx+426},
      ISBN = {978-0387-40080-8; 0-387-40080-X},
   MRCLASS = {62-02 (62-07 62H25 62H30)},
  MRNUMBER = {2168993},
}

@article {Yao2005,
    AUTHOR = {Yao, Fang and M\"uller, Hans-Georg and Wang, Jane-Ling},
     TITLE = {Functional data analysis for sparse longitudinal data},
   JOURNAL = {J. Amer. Statist. Assoc.},
  FJOURNAL = {Journal of the American Statistical Association},
    VOLUME = {100},
      YEAR = {2005},
    NUMBER = {470},
     PAGES = {577--590},
      ISSN = {0162-1459,1537-274X},
   MRCLASS = {62H25 (62G05)},
  MRNUMBER = {2160561},
MRREVIEWER = {M.\ Riedel},
       DOI = {10.1198/016214504000001745},
       URL = {https://doi.org/10.1198/016214504000001745},
}

@book {Horvath2012,
    AUTHOR = {Horv\'ath, Lajos and Kokoszka, Piotr},
     TITLE = {Inference for functional data with applications},
    SERIES = {Springer Series in Statistics},
 PUBLISHER = {Springer, New York},
      YEAR = {2012},
     PAGES = {xiv+422},
      ISBN = {978-1-4614-3654-6},
   MRCLASS = {62-02 (62G08 62H20 62H25 62J05 62M10 62M30)},
  MRNUMBER = {2920735},
MRREVIEWER = {Antonio\ Cuevas},
       DOI = {10.1007/978-1-4614-3655-3},
       URL = {https://doi.org/10.1007/978-1-4614-3655-3},
}

@book {Kokoszka2017,
    AUTHOR = {Kokoszka, Piotr and Reimherr, Matthew},
     TITLE = {Introduction to Functional Data Analysis},
    SERIES = {Texts in Statistical Science Series},
 PUBLISHER = {CRC Press, Boca Raton, FL},
      YEAR = {2017},
     PAGES = {xvi+290},
      ISBN = {978-1-4987-4634-2},
   MRCLASS = {62-01 (62Gxx 62H25 62Jxx 62M10 62M30)},
  MRNUMBER = {3793167},
MRREVIEWER = {David\ Benner\ Hitchcock},
}

@article {Golovkine2022,
	AUTHOR = {Golovkine, Steven and Klutchnikoff, Nicolas and Patilea,
	Valentin},
	TITLE = {Learning the smoothness of noisy curves with application to
	online curve estimation},
	JOURNAL = {Electron. J. Stat.},
	FJOURNAL = {Electronic Journal of Statistics},
	VOLUME = {16},
	YEAR = {2022},
	NUMBER = {1},
	PAGES = {1485--1560},
	MRNUMBER = {4390502},
	DOI = {10.1214/22-EJS1997},
}

@article{maissoro2024,
  title={Adaptive estimation for weakly dependent functional times series},
  author={Maissoro, Hassan and Patilea, Valentin and Vimond, Myriam},
  journal={Journal of Time Series Analysis},
  year={2025},
  publisher={Wiley Online Library},
  doi = {https://doi.org/10.1111/jtsa.70006}
}

@article{wang2023adaptive,
	title={Adaptive functional principal components analysis},
	author={Wang, Sunny G W and Patilea, Valentin and Klutchnikoff, Nicolas},
	fjournal = {Journal of the Royal Statistical Society: Series B},
  	journal={J. R. Stat. Soc. Ser. B. Stat. Methodol.},
	DOI={10.1093/jrsssb/qkae106},
        volume = {87},
        number = {3},
        pages = {603--631},
        year = {2025},
        month = {12},
}

@article {kassi2023,
    AUTHOR = {Kassi, Omar and Klutchnikoff, Nicolas and Patilea, Valentin},
     TITLE = {Learning the regularity of multivariate functional data},
   JOURNAL = {Electron. J. Stat.},
  FJOURNAL = {Electronic Journal of Statistics},
    VOLUME = {19},
      YEAR = {2025},
    NUMBER = {2},
     PAGES = {4174--4229},
      ISSN = {1935-7524},
   MRCLASS = {62G05 (60G22 62M09 62R10)},
  MRNUMBER = {4958400},
       DOI = {10.1214/25-ejs2433},
       URL = {https://doi-org.remotexs.ntu.edu.sg/10.1214/25-ejs2433},
}

@article {lepski-aniso-2015,
    AUTHOR = {Lepski, Oleg},
     TITLE = {Adaptive estimation over anisotropic functional classes via
              oracle approach},
   JOURNAL = {Ann. Statist.},
  FJOURNAL = {The Annals of Statistics},
    VOLUME = {43},
      YEAR = {2015},
    NUMBER = {3},
     PAGES = {1178--1242},
      ISSN = {0090-5364,2168-8966},
   MRCLASS = {62G05 (62G20)},
  MRNUMBER = {3346701},
MRREVIEWER = {Debdeep\ Pati},
       DOI = {10.1214/14-AOS1306},
       URL = {https://doi.org/10.1214/14-AOS1306},
}

@article {samarov-tsybakov-2004,
    AUTHOR = {Samarov, Alexander and Tsybakov, Alexandre},
     TITLE = {Nonparametric independent component analysis},
   JOURNAL = {Bernoulli},
  FJOURNAL = {Bernoulli. Official Journal of the Bernoulli Society for
              Mathematical Statistics and Probability},
    VOLUME = {10},
      YEAR = {2004},
    NUMBER = {4},
     PAGES = {565--582},
      ISSN = {1350-7265,1573-9759},
   MRCLASS = {62G07 (62G08 62G20)},
  MRNUMBER = {2076063},
MRREVIEWER = {Jack\ Cuzick},
       DOI = {10.3150/bj/1093265630},
       URL = {https://doi.org/10.3150/bj/1093265630},
}

@article {lepski-gilles-2020,
    AUTHOR = {Lepski, Oleg V. and Rebelles, Gilles},
     TITLE = {Structural adaptation in the density model},
   JOURNAL = {Math. Stat. Learn.},
  FJOURNAL = {Mathematical Statistics and Learning},
    VOLUME = {3},
      YEAR = {2020},
    NUMBER = {3-4},
     PAGES = {345--386},
      ISSN = {2520-2316,2520-2324},
   MRCLASS = {62G07 (62G05 62G20)},
  MRNUMBER = {4362042},
MRREVIEWER = {Paulo\ E.\ Oliveira},
       DOI = {10.4171/msl/24},
       URL = {https://doi.org/10.4171/msl/24},
}

@article{ammous-2024-dir,
	author = {Sinda Ammous and J{\'e}r{\^o}me Dedecker and C{\'e}line Duval},
	doi = {https://doi.org/10.1016/j.jmva.2024.105332},
	issn = {0047-259X},
	journal = {J. Multivariate Anal.},
	pages = {105332},
	title = {Adaptive directional estimator of the density in {$\mathbb R^d$} for independent and mixing sequences},
	url = {https://www.sciencedirect.com/science/article/pii/S0047259X24000393},
	volume = {203},
	year = {2024}
	}

@article {davies1999fractal,
    AUTHOR = {Davies, Steve and Hall, Peter},
     TITLE = {Fractal analysis of surface roughness by using spatial data},
      NOTE = {With discussion and a reply by the authors},
   JOURNAL = {J. R. Stat. Soc. Ser. B Stat. Methodol.},
  FJOURNAL = {Journal of the Royal Statistical Society. Series B.
              Statistical Methodology},
    VOLUME = {61},
      YEAR = {1999},
    NUMBER = {1},
     PAGES = {3--37},
      ISSN = {1369-7412,1467-9868},
   MRCLASS = {62H11 (62G05)},
  MRNUMBER = {1664088},
       DOI = {10.1111/1467-9868.00160},
       URL = {https://doi.org/10.1111/1467-9868.00160},
}

@article{bernardi2018,
	author = {Bernardi, Mara S. and Carey, Michelle and Ramsay, James O. and Sangalli, Laura M.},
	journal = {J. Multivariate Anal.},
	pages = {15--30},
	title = {Modeling spatial anisotropy via regression with partial differential regularization},
	volume = {167},
	year = {2018}
	}

@article{hsing2020,
author = {Jinqi Shen and Tailen Hsing},
title = {{Hurst function estimation}},
volume = {48},
JOURNAL = {Ann. Statist.},
fjournal = {The Annals of Statistics},
number = {2},
publisher = {Institute of Mathematical Statistics},
pages = {838--862},
year = {2020},
doi = {10.1214/19-AOS1825},
}

@article{herbin_06,
	author = {Herbin, Erick},
	doi = {10.1216/rmjm/1181069415},
	fjournal = {The Rocky Mountain Journal of Mathematics},
	issn = {0035-7596},
	journal = {Rocky Mountain J. Math.},
	mrclass = {62G05 (60G15)},
	mrnumber = {2274895},
	mrreviewer = {Ciprian A. Tudor},
	number = {4},
	pages = {1249--1284},
	title = {From {$N$} parameter fractional {B}rownian motions to {$N$} parameter multifractional {B}rownian motions},
	url = {https://doi.org/10.1216/rmjm/1181069415},
	volume = {36},
	year = {2006}
	}

@incollection{fan2016multivariate,
	author = {Fan, Yangin and Guerre, Emmanuel},
	booktitle = {Essays in Honor of Aman Ullah},
	pages = {489--537},
	publisher = {Emerald Group Publishing Limited},
	title = {Multivariate local polynomial estimators: uniform boundary properties and asymptotic linear representation},
	volume = {36},
	year = {2016}
}

@article{woodchan94,
	author = {Wood, Andrew T. A. and Chan, Grace},
	doi = {10.2307/1390903},
	fjournal = {Journal of Computational and Graphical Statistics},
	issn = {1061-8600,1537-2715},
	journal = {J. Comput. Graph. Statist.},
	mrclass = {65C20 (60G10)},
	mrnumber = {1323050},
	number = {4},
	pages = {409--432},
	title = {Simulation of stationary {G}aussian processes in {$[0,1]^d$}},
	url = {https://doi.org/10.2307/1390903},
	volume = {3},
	year = {1994}
	}

@article{stein2002,
	author = {Stein, Michael L.},
	doi = {10.1198/106186002466},
	fjournal = {Journal of Computational and Graphical Statistics},
	issn = {1061-8600,1537-2715},
	journal = {J. Comput. Graph. Statist.},
	mrclass = {60G15 (65D18)},
	mrnumber = {1938447},
	number = {3},
	pages = {587--599},
	title = {Fast and exact simulation of fractional {B}rownian surfaces},
	url = {https://doi.org/10.1198/106186002466},
	volume = {11},
	year = {2002}
	}

@article{coeurjolly2018,
	author = {Coeurjolly, Jean-Francois and Porcu, Emilio},
	doi = {10.1080/10618600.2017.1385468},
	fjournal = {Journal of Computational and Graphical Statistics},
	issn = {1061-8600,1537-2715},
	journal = {J. Comput. Graph. Statist.},
	mrclass = {60G22 (60G15 62M09)},
	mrnumber = {3816264},
	number = {2},
	pages = {278--290},
	title = {Fast and exact simulation of complex-valued stationary {G}aussian processes through embedding circulant matrix},
	url = {https://doi.org/10.1080/10618600.2017.1385468},
	volume = {27},
	year = {2018}
	}

@mastersthesis{dieker2004,
  author  = {Dieker, Ton},
  title   = {Simulation of fractional brownian motion},
  school  = {University of Twente},
  year    = {2004},
  type    = {Master’s thesis}
}

@article {matheron1973,
    AUTHOR = {Matheron, G.},
     TITLE = {The intrinsic random functions and their applications},
   JOURNAL = {Advances in Appl. Probability},
  FJOURNAL = {Advances in Applied Probability},
    VOLUME = {5},
      YEAR = {1973},
     PAGES = {439--468},
      ISSN = {0001-8678,1475-6064},
   MRCLASS = {60G20 (62M10)},
  MRNUMBER = {356209},
MRREVIEWER = {R.\ Jajte},
       DOI = {10.2307/1425829},
       URL = {https://doi.org/10.2307/1425829},
}

@article {richards2015,
    AUTHOR = {Bierm\'e, Hermine and Moisan, Lionel and Richard,
              Fr\'ed\'eric},
     TITLE = {A turning-band method for the simulation of anisotropic
              fractional {B}rownian fields},
   JOURNAL = {J. Comput. Graph. Statist.},
  FJOURNAL = {Journal of Computational and Graphical Statistics},
    VOLUME = {24},
      YEAR = {2015},
    NUMBER = {3},
     PAGES = {885--904},
      ISSN = {1061-8600,1537-2715},
   MRCLASS = {60G60 (60G18 62M40)},
  MRNUMBER = {3397238},
       DOI = {10.1080/10618600.2014.946603},
       URL = {https://doi.org/10.1080/10618600.2014.946603},
}

@article {ramsay2002,
    AUTHOR = {Ramsay, Tim},
     TITLE = {Spline smoothing over difficult regions},
   JOURNAL = {J. R. Stat. Soc. Ser. B Stat. Methodol.},
  FJOURNAL = {Journal of the Royal Statistical Society. Series B.
              Statistical Methodology},
    VOLUME = {64},
      YEAR = {2002},
    NUMBER = {2},
     PAGES = {307--319},
      ISSN = {1369-7412,1467-9868},
   MRCLASS = {62H12},
  MRNUMBER = {1904707},
       DOI = {10.1111/1467-9868.00339},
       URL = {https://doi.org/10.1111/1467-9868.00339},
}

@article {sangalli2013,
    AUTHOR = {Sangalli, Laura M. and Ramsay, James O. and Ramsay, Timothy
              O.},
     TITLE = {Spatial spline regression models},
   JOURNAL = {J. R. Stat. Soc. Ser. B. Stat. Methodol.},
  FJOURNAL = {Journal of the Royal Statistical Society. Series B.
              Statistical Methodology},
    VOLUME = {75},
      YEAR = {2013},
    NUMBER = {4},
     PAGES = {681--703},
      ISSN = {1369-7412,1467-9868},
   MRCLASS = {62H11 (62G08)},
  MRNUMBER = {3091654},
       DOI = {10.1111/rssb.12009},
       URL = {https://doi.org/10.1111/rssb.12009},
}

@article {azzim2015,
    AUTHOR = {Azzimonti, Laura and Sangalli, Laura M. and Secchi, Piercesare
              and Domanin, Maurizio and Nobile, Fabio},
     TITLE = {Blood flow velocity field estimation via spatial regression
              with {PDE} penalization},
   JOURNAL = {J. Amer. Statist. Assoc.},
  FJOURNAL = {Journal of the American Statistical Association},
    VOLUME = {110},
      YEAR = {2015},
    NUMBER = {511},
     PAGES = {1057--1071},
      ISSN = {0162-1459,1537-274X},
   MRCLASS = {62M40 (62G08 62H11 62P10)},
  MRNUMBER = {3420684},
       DOI = {10.1080/01621459.2014.946036},
       URL = {https://doi.org/10.1080/01621459.2014.946036},
}

@article{jainhong1997,
	author = {Jain, A. and Lin Hong and Bolle, R.},
	doi = {10.1109/34.587996},
        journal = {IEEE Trans. Pattern Anal. Mach. Intell.},
	fjournal = {IEEE Transactions on Pattern Analysis and Machine Intelligence},
	number = {4},
	pages = {302--314},
	title = {On-line fingerprint verification},
	volume = {19},
	year = {1997}
	}

@article {jiang2005,
    AUTHOR = {Jiang, Xudong},
     TITLE = {On orientation and anisotropy estimation for online
              fingerprint authentication},
   JOURNAL = {IEEE Trans. Signal Process.},
  FJOURNAL = {IEEE Transactions on Signal Processing},
    VOLUME = {53},
      YEAR = {2005},
    NUMBER = {10},
     PAGES = {4038--4049},
      ISSN = {1053-587X,1941-0476},
   MRCLASS = {99-01},
  MRNUMBER = {2246664},
       DOI = {10.1109/TSP.2005.855417},
       URL = {https://doi.org/10.1109/TSP.2005.855417},
}

@article{GERMAIN2003,
	author = {C. Germain and J.P. {Da Costa} and O. Lavialle and P. Baylou},
	doi = {https://doi.org/10.1016/S0165-1684(03)00064-1},
	journal = {Signal Processing},
	number = {7},
	pages = {1487--1503},
	title = {Multiscale estimation of vector field anisotropy application to texture characterization},
	volume = {83},
	year = {2003}
	}

@article {richard2016test,
    AUTHOR = {Richard, Fr\'ed\'eric J. P.},
     TITLE = {Tests of isotropy for rough textures of trended images},
   JOURNAL = {Statist. Sinica},
  FJOURNAL = {Statistica Sinica},
    VOLUME = {26},
      YEAR = {2016},
    NUMBER = {3},
     PAGES = {1279--1304},
      ISSN = {1017-0405,1996-8507},
   MRCLASS = {94A08 (60G17 62F12 62H15 62M40)},
  MRNUMBER = {3559953},
MRREVIEWER = {Rachid\ Jennane},
}

@article {wood2008,
    AUTHOR = {Wood, Simon N. and Bravington, Mark V. and Hedley, Sharon L.},
     TITLE = {Soap film smoothing},
   JOURNAL = {J. R. Stat. Soc. Ser. B Stat. Methodol.},
  FJOURNAL = {Journal of the Royal Statistical Society. Series B.
              Statistical Methodology},
    VOLUME = {70},
      YEAR = {2008},
    NUMBER = {5},
     PAGES = {931--955},
      ISSN = {1369-7412,1467-9868},
   MRCLASS = {99-01},
  MRNUMBER = {2530324},
       DOI = {10.1111/j.1467-9868.2008.00665.x},
       URL = {https://doi.org/10.1111/j.1467-9868.2008.00665.x},
}

@misc{prism2025,
  author        = {PRISM Climate Group, Oregon State University},
  title         = {PRISM climate data},
  howpublished  = {\url{https://prism.oregonstate.edu/}},
  note          = {Accessed: 2025-06-19},
  year          = {2025}
}

@techreport{daly-2013,
	author = {Daly, Christopher and Bryant, Kirk and Oregon State University and USDA Natural Resources Conservation Service},
	month = {6},
	title = {{The PRISM climate and weather system – an introduction.}},
	year = {2013},
	url = {https://prism.oregonstate.edu/documents/PRISM_history_jun2013.pdf},
}

@Manual{hart2015prism,
  title        = {PRISM: download data from the Oregon state prism climate project},
  author       = {Edmund M. Hart and Kendon Bell},
  year         = {2015},
  note         = {R package version 0.2.3},
  url          = {https://cran.r-project.org/package=prism},
  doi          = {10.32614/CRAN.package.prism},
}

@article{oliveira2022,
  author = {Oliveira, A.L.G. and Lima, J.P. and Brasco, T.L. and Amaral, L.R.},
  title = {The importance of modeling the effects of trend and anisotropy on soil fertility maps},
  journal = {Computers and Electronics in Agriculture},
  volume = {196},
  pages = {106877},
  year = {2022},
  doi = {10.1016/j.compag.2022.106877}
}

@article{friedland2016,
  author = {Friedland, C.J. and Joyner, T.A. and Massarra, C. and Rohli, R.V. and others},
  title = {Isotropic and anisotropic kriging approaches for interpolating surface-level wind speeds across large, geographically diverse regions},
  journal = {Geomatics, Natural Hazards and Risk},
  volume = {8},
  number = {2},
  pages = {1--18},
  year = {2016},
  doi = {10.1080/19475705.2016.1185749}
}

@article{niemi2015,
  author = {Niemi, T.J. and Guillaume, J.H.A. and Kokkonen, T. and Hoang, T.M.T. and Seed, A.W.},
  title = {Role of spatial anisotropy in design storm generation: experiment and interpretation},
  journal = {Water Resources Research},
  volume = {51},
  number = {1},
  pages = {1--21},
  year = {2015},
  doi = {10.1002/2015WR017521}
}

@inproceedings{wang2005,
  author    = {Yingzi Wang and Jie Hu and Hong Yan},
  title     = {A gradient-based weighted averaging method for estimation of fingerprint orientation fields},
  booktitle = {Proceedings of the Digital Image Computing: Techniques and Applications (DICTA)},
  year      = {2005},
  pages     = {191--198}
}

@article{greenberg2002,
  author  = {Shlomo Greenberg and Michael Aladjem and Daniel Kogan},
  title   = {Fingerprint image enhancement using filtering techniques},
  journal = {Real-Time Imaging},
  volume  = {8},
  number  = {3},
  year    = {2002},
  pages   = {227--236},
  doi     = {10.1006/rtim.2002.0279}
}

@article{somvanshi2012,
  author  = {Prachi Somvanshi and Mayuresh Rane},
  title   = {Survey of palmprint recognition},
  journal = {International Journal of Scientific \& Engineering Research},
  volume  = {3},
  number  = {2},
  year    = {2012},
  pages   = {1--5}
}

@article{kassiwangsupp25,
  author  = {Omar Kassi and Sunny G W Wang},
  title   = {Supplement to "Structural adaptation and rate accelerated estimation in bivariate functional data"},
  JOURNAL = {Bernoulli},
  FJOURNAL = {Bernoulli. Official Journal of the Bernoulli Society for
              Mathematical Statistics and Probability},
  year = {2025}

}

@article {amorino2021,
    AUTHOR = {Amorino, Chiara},
     TITLE = {Rate of estimation for the stationary distribution of
              jump-processes over anisotropic {H}older classes},
   JOURNAL = {Electron. J. Stat.},
  FJOURNAL = {Electronic Journal of Statistics},
    VOLUME = {15},
      YEAR = {2021},
    NUMBER = {2},
     PAGES = {5067--5116},
      ISSN = {1935-7524},
   MRCLASS = {62G07 (62G20)},
  MRNUMBER = {4347384},
MRREVIEWER = {Junlian\ Xu},
       DOI = {10.1214/21-ejs1913},
       URL = {https://doi-org.remotexs.ntu.edu.sg/10.1214/21-ejs1913},
}

@article {bertin2004,
    AUTHOR = {Bertin, Karine},
     TITLE = {Asymptotically exact minimax estimation in sup-norm for
              anisotropic {H}\"older classes},
   JOURNAL = {Bernoulli},
  FJOURNAL = {Bernoulli. Official Journal of the Bernoulli Society for
              Mathematical Statistics and Probability},
    VOLUME = {10},
      YEAR = {2004},
    NUMBER = {5},
     PAGES = {873--888},
      ISSN = {1350-7265,1573-9759},
   MRCLASS = {62C20 (46N30 62F12)},
  MRNUMBER = {2093615},
MRREVIEWER = {B.\ M. Golam Kibria},
       DOI = {10.3150/bj/1099579160},
       URL = {https://doi-org.remotexs.ntu.edu.sg/10.3150/bj/1099579160},
}

@article {rice2001,
    AUTHOR = {Rice, John A. and Wu, Colin O.},
     TITLE = {Nonparametric mixed effects models for unequally sampled noisy
              curves},
   JOURNAL = {Biometrics},
  FJOURNAL = {Biometrics. Journal of the International Biometric Society},
    VOLUME = {57},
      YEAR = {2001},
    NUMBER = {1},
     PAGES = {253--259},
      ISSN = {0006-341X,1541-0420},
   MRCLASS = {99-01},
  MRNUMBER = {1833314},
       DOI = {10.1111/j.0006-341X.2001.00253.x},
       URL = {https://doi-org.remotexs.ntu.edu.sg/10.1111/j.0006-341X.2001.00253.x},
}


\clearpage
\includepdf[pages={-}, fitpaper=true]{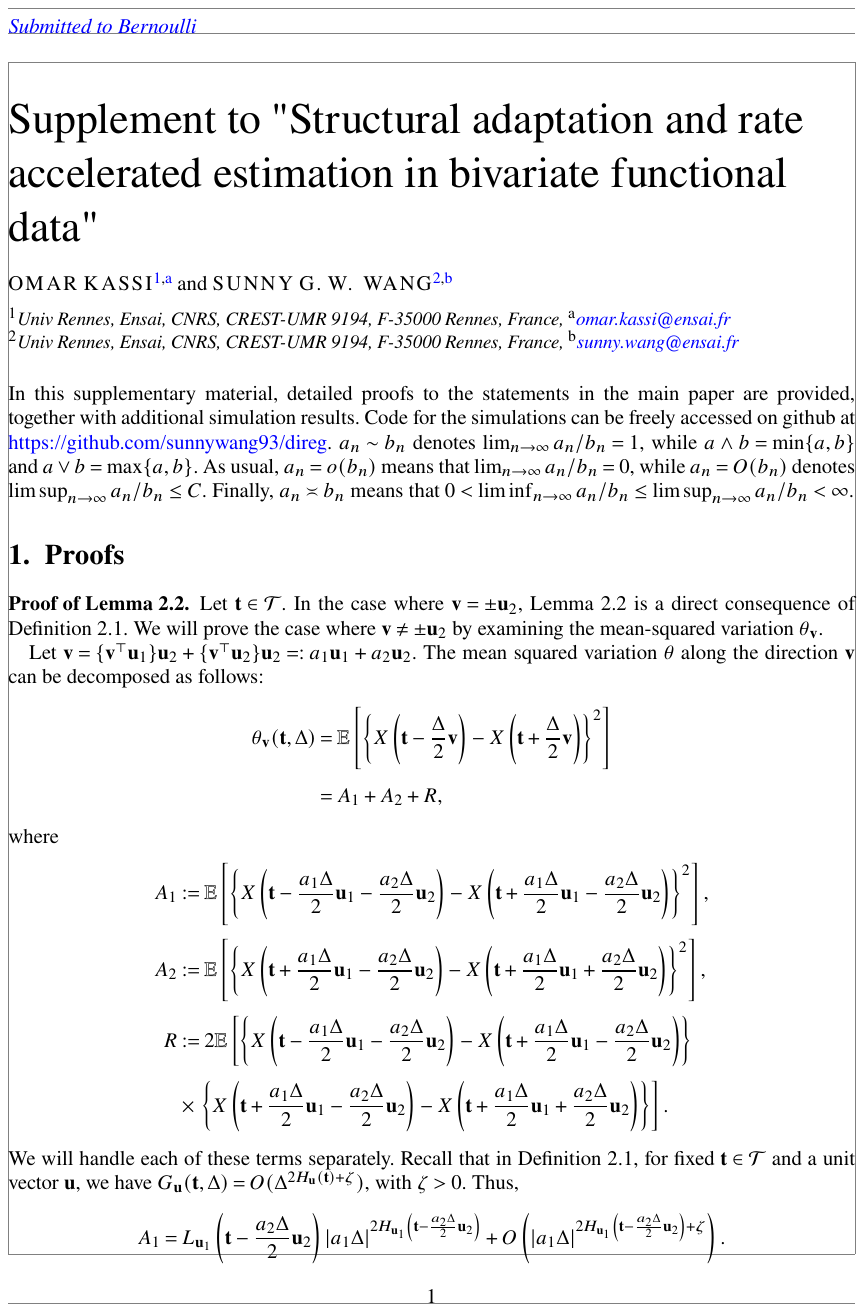}

\end{document}